\documentclass[superscriptaddress,
 reprint,
 amssymb, amsmath,
 aip, cha 
]{revtex4-1}
\usepackage[utf8]{inputenc}
\usepackage[T1]{fontenc}
\usepackage{amsmath}
\usepackage{amsfonts}
\usepackage{amssymb}
\usepackage{graphicx}
\usepackage{bm}
\usepackage{grffile} 
\usepackage[dvipsnames]{xcolor}

\newcommand*{\citen}[1]{%
  \begingroup
    \romannumeral-`\x 
    \setcitestyle{numbers}%
    \cite{#1}%
  \endgroup   
}

\begin{document}


\title{Chaos in networks of coupled oscillators with multimodal natural frequency distributions}

\author{Lachlan~D. Smith}
 \email{lachlan.smith@sydney.edu.au}
 \affiliation{\mbox{School of Mathematics and Statistics, The University of Sydney, Sydney, NSW 2006, Australia}}
\author{Georg~A. Gottwald}
 \email{georg.gottwald@sydney.edu.au}
 \affiliation{\mbox{School of Mathematics and Statistics, The University of Sydney, Sydney, NSW 2006, Australia}}

\date{\today}

\begin{abstract}
We explore chaos in the Kuramoto model with multimodal distributions of the natural frequencies of oscillators and provide a comprehensive description under what conditions chaos occurs. For a natural frequency distribution with $M$ peaks it is typical that there is a range of coupling strengths such that oscillators belonging to each peak form a synchronized cluster, but the clusters do not globally synchronize. We use collective coordinates to describe the inter- and intra-cluster dynamics, which reduces the Kuramoto model to $2M-1$ degrees of freedom. We show that under some assumptions, there is a time-scale splitting between the slow intracluster dynamics and fast intercluster dynamics, which reduces the collective coordinate model to an $M-1$ degree of freedom rescaled Kuramoto model. Therefore, four or more clusters are required to yield the three degrees of freedom necessary for chaos. However, the time-scale splitting breaks down if a cluster intermittently desynchronizes. We show that this intermittent desynchronization provides a mechanism for chaos for trimodal natural frequency distributions. In addition, we use collective coordinates to show analytically that chaos cannot occur for bimodal frequency distributions, even if they are asymmetric and if intermittent desynchronization occurs.
\end{abstract}

\maketitle

\begin{quotation}
Synchronization of coupled oscillators occurs in many natural processes and engineering applications. The dynamics of the globally synchronized state is regular and the phases typically rotate with a constant mean frequency. In the case of multimodal distributions of natural frequencies of the oscillators, one can observe more complex dynamics including chaos. Under which conditions the synchronized state may exhibit chaos has not been fully addressed. Distinct peaks in a multimodal natural frequency distribution correspond to synchronized clusters for a range of coupling strengths and network parameters. We study the intercluster and intracluster dynamics using a collective coordinate approach, which reduces the dimension of the full Kuramoto model to a small number of active degrees of freedom. We find necessary conditions for chaos to occur. In particular, at least four peaks in the natural frequency distribution are required to produce phase chaos, and chaos can also occur for three peaks via intermittent desynchronization of clusters.
\end{quotation}

\section{Introduction}

Synchronization in networks of coupled oscillators occurs in many natural systems, including the activity of the brain \cite{SheebaEtAl08, BhowmikShanahan12} and synchronous firefly flashing, \cite{MirolloStrogatz90} as well as many engineering applications, such as power grids, \cite{FilatrellaEtAl08} and Josephson junction arrays.  \cite{WatanabeStrogatz94, WiesenfeldEtAl98}

In typical models of synchronization, the dynamics is either incoherent, partially synchronized, or fully synchronized. In the case of a unimodal frequency distribution, the dynamics transitions upon increasing the coupling strength from the incoherent state at low coupling strength, to a partially synchronized state where a collection of oscillators synchronize (those with native frequency closest to the mean frequency), to the fully synchronized state at high coupling strengths. For multimodal frequency distributions, however, several synchronized clusters may emerge in the partially-synchronized regime. That is, there are clusters of oscillators that remain synchronized within themselves, but the oscillators do not form a single synchronized cluster. These clusters may have complex interactions, both inter-cluster and intra-cluster, producing complex dynamics, including chaos.

Chaos in coupled oscillator networks has been previously studied. For the Kuramoto model, \cite{Kuramoto84, Strogatz00, PikovskyEtAl01, AcebronEtAl05, OsipovEtAl07, ArenasEtAl08, DorflerBullo14, RodriguesEtAl16} which is the model focused on here, chaos has been observed in the incoherent state, termed phase chaos, \cite{PopovychEtAl05, MaistrenkoEtAl05, MiritelloEtAl09} provided there are at least four oscillators. This type of phase chaos occurs at the microscopic level and is associated with the chaotic dynamics of individual phase oscillators. For such microscopic phase chaos, the Lyapunov exponent was found to scale inversely proportionally to the number of oscillators. \cite{CarluEtAl18} In particular, this implies that in the thermodynamic limit of infinitely many oscillators the Lyapunov exponent is zero, i.e., no chaos. Here we focus on collective chaotic behavior of synchronized subpopulations of phase oscillators. Such collective chaos has been studied for systems with symmetric bimodal natural frequency distributions which were subjected to a time-periodic coupling strength, \cite{SoBarreto11} or for different inter- and intra-cluster coupling strengths as well as a phase lag. \cite{BickEtAl18} However, for the classical Kuramoto model, it has been shown that in the thermodynamic limit with bimodal natural frequency distributions chaos is impossible. \cite{SoBarreto11} For trimodal frequency distributions, which yield three synchronized subpopulations, chaos has been observed for superposed Lorentzian natural frequency distributions, but only in the partially synchronized state, which involves microscopic chaos of incoherent oscillators. \cite{ChengGuo17}

Here we present and analytically study generic situations of collective chaos in which the dynamics of synchronized subpopulations of coupled oscillators, termed clusters, can be chaotic.  We distinguish between two types of chaotic dynamics, one akin to phase chaos and the other due to intermittent desynchronization. Here we refer to collective phase chaos when each of the synchronized clusters preserves their shape while the phases of the clusters show chaotic behavior. In this case, the possibility of chaos is determined by the number of synchronized clusters, which determines the number of active degrees of freedom. We shall see that to obtain phase chaos at least four synchronized clusters are necessary. This is analogous to needing at least four oscillators to generate microscopic phase chaos in the incoherent state of the Kuramoto model. \cite{PopovychEtAl05, MaistrenkoEtAl05, MiritelloEtAl09}

A different type of chaos is observed when clusters intermittently desynchronize through their mutual interactions. In this case, as we will show, chaos may occur even for trimodal natural frequency distributions.

The key underlying reason for both types of chaos is that chaos can only occur when there are at least three degrees of freedom. Each synchronized cluster can be characterized by a time-varying shape variable and a mean phase variable, which are the active degrees of freedom, and the interaction of these collective coordinates can lead to chaos. We reduce the full Kuramoto model to the evolution equations for these collective coordinates. \cite{Gottwald15, Gottwald17, HancockGottwald18} We demonstrate a time-scale splitting between the (slow) shape and the (fast) phase variables, that enables further reduction. Under this reduction, the full Kuramoto model with $M$ clusters reduces to a renormalized Kuramoto model with $M$ oscillators, which has $M-1$ degrees of freedom, implying that $M\geq 4$ is necessary for phase chaos to occur. However, when a cluster intermittently desynchronizes, the time-scale splitting is invalid, yielding additional active degrees of freedom, and the potential for chaos with three clusters.

The paper is organized as follows; in \S\ref{sec:the_model} we describe the Kuramoto model. Then in \S\ref{sec:collective_coordinates} we present the collective coordinate ansatz and derive the evolution equations for the collective coordinates. In \S\ref{sec:four_clusters} we show that phase chaos occurs for four clusters, and that there is quantitative agreement between the leading Lyapunov exponent for the full Kuramoto model and the collective coordinate reduction. In \S\ref{sec:three_clusters} we show that chaos can occur for three clusters via intermittent desynchronization of a cluster, and provide a detailed description of this mechanism. Again, there is quantitative agreement between the leading Lyapunov exponent for the full Kuramoto model and the collective coordinate reduction. In \S\ref{sec:two_clusters} we show that chaos is not possible for two clusters in the thermodynamic limit of infinitely many oscillators. Lastly, in \S\ref{sec:conclusions} we summarize our results and provide an outlook for future studies.

\section{The model} \label{sec:the_model}

The Kuramoto model has been widely used to model networks of coupled oscillators, \cite{Kuramoto84, Strogatz00, PikovskyEtAl01, AcebronEtAl05, OsipovEtAl07, ArenasEtAl08, DorflerBullo14, RodriguesEtAl16} in large part due to its analytical tractability. For a network of $N$ coupled oscillators, each with phase $\phi_i$, the dynamics is given by
\begin{equation} \label{eq:full_KM}
\dot{\phi_i} = \omega_i + \frac{K}{N} \sum_{j=1}^N A_{ij} \sin(\phi_j - \phi_i),
\end{equation}
where the natural frequencies $\omega_i$ are drawn from a distribution $g(\omega)$, $A$ is the adjacency matrix of the network, i.e., $A_{ij}=1$ if nodes $i$ and $j$ are connected, otherwise $A_{ij}=0$, and $K$ is the coupling strength. We shall restrict our study of collective chaos to an all-to-all coupling topology with $A_{ij}=1-\delta_{ij}$. For the exposition of the model reduction technique presented in \S\ref{sec:collective_coordinates}, however, we choose to present the Kuramoto model (\ref{eq:full_KM}) with a general topology. It is widely known that if the coupling strength is sufficiently large, then the oscillators spontaneously synchronize, all oscillating at the same frequency, even though their natural frequencies are different. Furthermore, below the global synchronization threshold, synchronized clusters may emerge due to either clusters in the network topology, or distinct modes in the natural frequency distribution, or both. 

\begin{figure}[tbp]
\centering
\includegraphics[width=0.8\columnwidth]{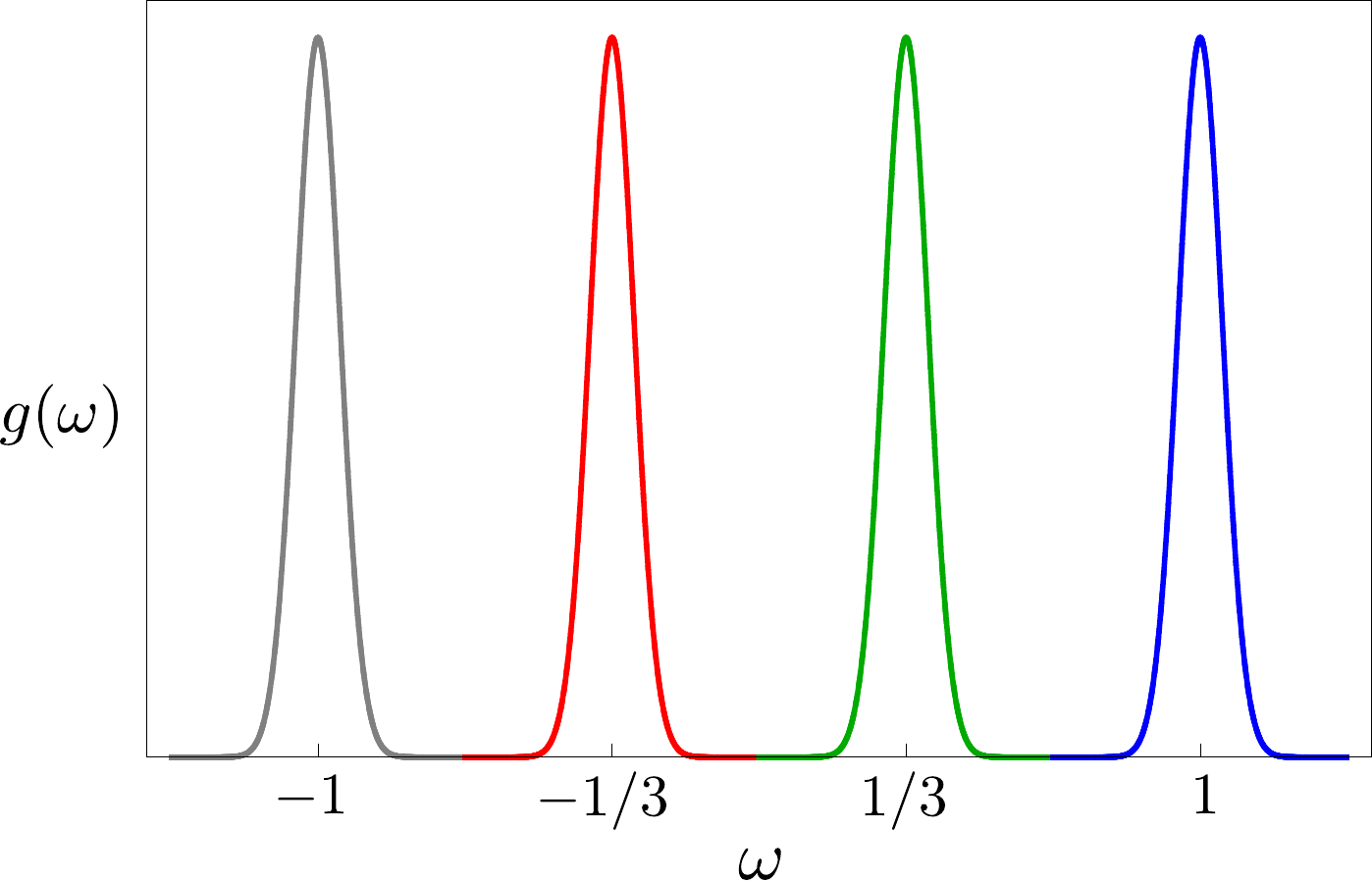}
\caption{A multimodal natural frequency distribution of the form (\ref{eq:multimodal_distribution}) with four peaks and equal weights, $\gamma_m = 1/4$. The means $\Omega_m$ are equally spaced between $-1$ and $1$ and the standard deviations are all equal with $\sigma_m=0.05$.}
\label{fig:quadmodal_distro}
\end{figure}

We consider multimodal natural frequency distributions $g(\omega)$ of the form
\begin{equation} \label{eq:multimodal_distribution}
g(\omega) = \sum_{m=1}^M \gamma_m g_m(\omega; \Omega_m, \sigma_m^2),
\end{equation}
such that each $g_m$ is a normal distribution with mean $\Omega_m$ and variance $\sigma_m^2$, and the weights $0\leq\gamma_m\leq 1$ satisfy $\sum \gamma_m = 1$. In particular, we primarily consider the case of well-separated peaks, such as the the example shown in Fig.~\ref{fig:quadmodal_distro}. The distribution (\ref{eq:multimodal_distribution}) has $M$ peaks, which typically correspond to $M$ clusters of synchronized oscillators for a range of coupling strengths. Note that the Kuramoto model is invariant under uniform phase shifts. Therefore, we may assume without loss of generality that the mean natural frequency is zero, i.e., $\sum_m \gamma_m \Omega_m = 0$.

A characterization of the state of the system is the instantaneous order parameter $r(t)$ which is defined as
\begin{equation} \nonumber
r(t) e^{i \psi(t)} = \frac{1}{N} \sum_{j=1}^N e^{i \phi_j (t)},
\end{equation}
and describes the mean position of all oscillators in the complex plane. The long term dynamics can be characterized by the time-averaged order parameter
\begin{equation} \nonumber
\bar{r} = \frac{1}{T} \int_{t_0}^{t_0+T} r(t) dt ,
\end{equation}
which is independent of $t_0$ and $T$ for\ sufficiently long transient times $t_0$ and averaging times $T$. If $\bar{r}$ is close to $1$, the oscillators are globally synchronized. If $\bar{r} \approx 1/\sqrt{N}$, the oscillators are in the incoherent state. In addition, for cases with multiple synchronized clusters, we can define analogous instantaneous and time-averaged order parameters for each cluster. For example, the instantaneous order parameter for the $m$-th cluster is
\begin{equation} \nonumber
r_m(t) e^{i \psi_m(t)} = \frac{1}{N_m} \sum_{j \in \mathcal{C}_m} e^{i \phi_j (t)},
\end{equation}
where $\mathcal{C}_m$ is the set of oscillators in cluster $m$, and $N_m$ is the number of oscillators in $\mathcal{C}_m$. 

\begin{figure}[tbp]
\centering
\includegraphics[width=\columnwidth]{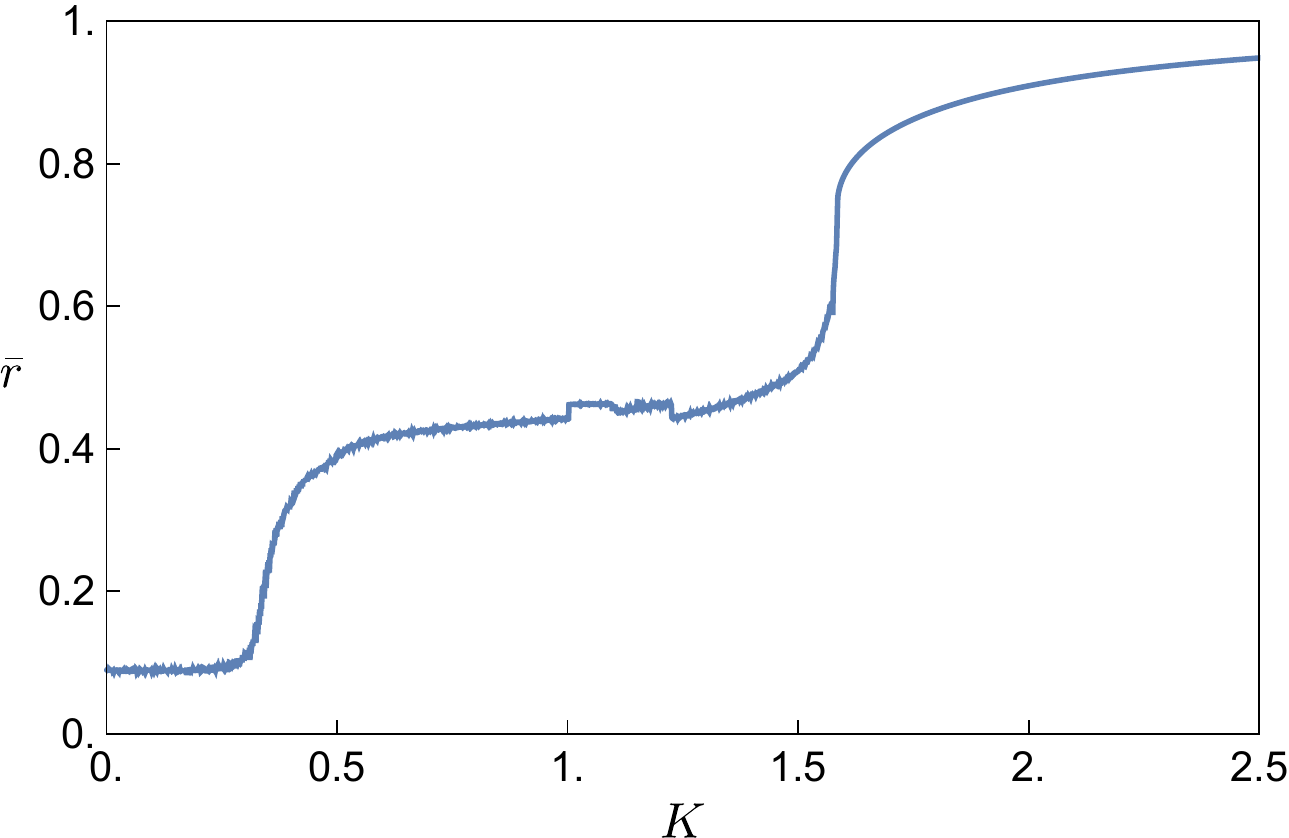}
\caption{Time averaged order parameter $\bar{r}$ for the multimodal natural frequency distribution shown in Fig.~\ref{fig:quadmodal_distro} over a range of coupling strengths $K$ for the Kuramoto model (\ref{eq:full_KM}) with $N=100$ oscillators.}
\label{fig:four_cluster_rbar_fixed_sigma}
\end{figure}

Consider, for example, the frequency distribution shown in Fig.~\ref{fig:quadmodal_distro}, with four peaks. The modes $\Omega_m$ are equally spaced between $-1$ and $1$, and the standard deviations are all the same with $\sigma_m=0.05$. For $N=100$ equiprobably\footnote{We find the values $\omega_i$, $i=1,\dots,N$, such that $G(\omega_i) = (i+1)/(N+1)$, where $G(\omega)$ is the cumulative density function of the natural frequencies. By sampling equiprobably we eliminate the need to average over many randomly drawn realizations.} drawn oscillators from this distribution, the time-averaged order parameter, $\bar{r}$, is shown for $0<K<2.5$ in Fig.~\ref{fig:four_cluster_rbar_fixed_sigma}. For $K<0.3$, the oscillators are incoherent, and $\bar{r}$ is of the order $1/\sqrt{N}$. For $K>1.6$, the oscillators globally synchronize, forming a single cluster, and $\bar{r}\approx 1$. For intermediate values, i.e., $0.3<K<1.6$, the oscillators corresponding to each peak in $g(\omega)$ synchronize to form a cluster, but they do not globally synchronize, resulting in $\bar{r} \approx 0.45$. In this study we are mostly interested in these intermediate values, where there can be complex interactions within and in between clusters. Note that $\bar{r}$ exhibits unusual non-monotonic behavior around $1<K<1.25$, which, as we shall see, is the region where chaotic dynamics occurs. 

\begin{figure}[tbp]
\centering
\includegraphics[width=\columnwidth]{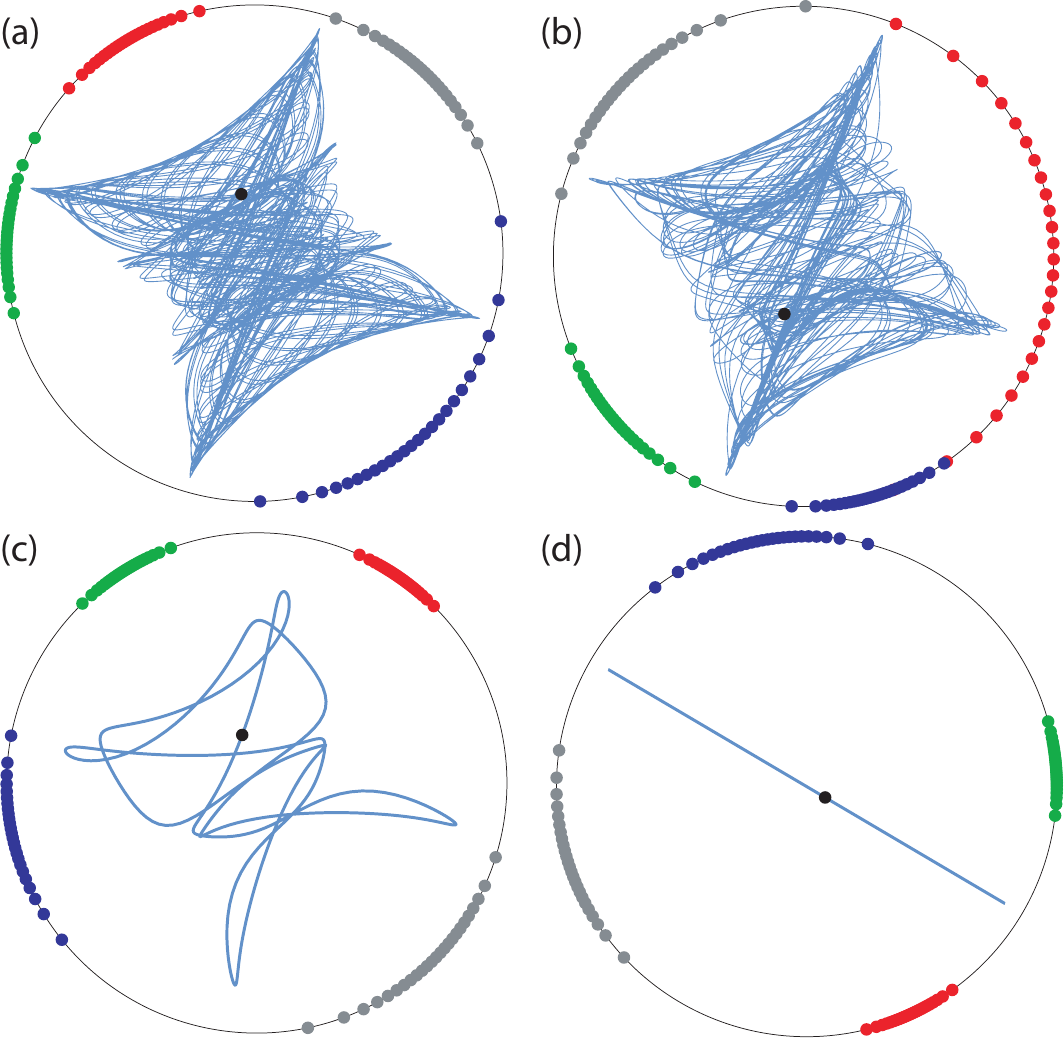}
\caption{Snapshots of oscillators in the complex plane, and the trajectories of the complex order parameter $r(t) e^{i\psi(t)}$ after a transient integration time for the Kuramoto model (\ref{eq:full_KM}) with $N=100$ oscillators with natural frequency distribution shown in Fig.~\ref{fig:quadmodal_distro}. (a)~For $K=0.9$ the dynamics are quasiperiodic. (Multimedia view) (b)~For $K=1.2$ the dynamics are chaotic. (Multimedia view) (c)~For $K=1.22$ the dynamics are periodic. (Multimedia view) (d)~For $K=1.3$ the dynamics are periodic, and the complex order parameter is confined to a symmetric invariant manifold (a straight line). (Multimedia view)}
\label{fig:complex_order_param_trajectories}
\end{figure}

Synchronization of clusters is shown by the snapshots of oscillators in the complex plane in Fig.~\ref{fig:complex_order_param_trajectories} for four different values of $K$. The oscillators of each color (corresponding to the same colored peak in Fig.~\ref{fig:quadmodal_distro}) are synchronized, but there are clearly four distinct clusters. These clusters have both their own internal dynamics and interact with other clusters. For $K=0.9$, the dynamics is quasiperiodic, demonstrated by the trajectory of the complex order parameter $r(t)e^{i\psi(t)}$ shown as the blue curve inside the circle in Fig.~\ref{fig:complex_order_param_trajectories}(a) (Multimedia view). Increasing $K$, at a critical coupling strength $K_c$ the dynamics becomes chaotic. For example, with $K=1.2$, shown in Fig.~\ref{fig:complex_order_param_trajectories}(b) (Multimedia view), the dynamics is chaotic (which is confirmed by computing the leading Lyapunov exponent, $\lambda = 6.18 \times 10^{-2}$). The dynamics then becomes regular again, for example with $K=1.22$ and $K=1.3$ the trajectory of the complex order parameter is periodic (cf. Fig.~\ref{fig:complex_order_param_trajectories}(c) (Multimedia view) and Fig.~\ref{fig:complex_order_param_trajectories}(d) (Multimedia view), respectively). For $K=1.3$, the trajectory is confined to a straight line due to the existence of an attracting symmetric manifold. Four cluster cases such as these will be discussed in more detail in \S\ref{sec:four_clusters}.

\section{Model reduction via collective coordinates} \label{sec:collective_coordinates}

Since we are primarily interested in the macroscopic inter- and intracluster dynamics, we use model reduction to reduce the high dimensional full Kuramoto model (\ref{eq:full_KM}) to a small number of active degrees of freedom. One frequently used method is the Ott-Antonsen approach, \cite{OttAntonson08} which assumes infinitely many oscillators. Recently, an alternative approach for model reduction has been proposed, termed collective coordinate reduction,\cite{Gottwald15, Gottwald17, HancockGottwald18} which can be readily applied to finite networks of coupled oscillators.

The idea of the collective coordinate reduction\cite{Gottwald15, Gottwald17, HancockGottwald18} is to express the $N$-dimensional phase vector $\bm{\phi}$ as a linear combination of a small number of dynamically relevant modes. Intuitively, the reduction is motivated by the fact that synchronization is characterized by oscillators forming a collective entity which is described by its mean phase and its shape. The time-varying coefficients of the linear combination are coined collective coordinates, and encode the temporal evolution of the modes. Identification of the relevant modes is situation-dependent. In the case of a single synchronized cluster of oscillators, where the global phase is not relevant, a single mode $\bm{\Phi}$ describing the shape suffices, and we approximate $\bm{\phi}(t) \approx \alpha(t) \bm{\Phi}$. When multiple clusters interact, phase variables need to be accounted for. We will denote the shape modes by $\bm{\Phi}^{(m)}$ and the phase modes by $\bm{1}_{N_m}$ (the vector consisting of $N_m$ $1$'s, where $N_m$ is the size of the $m$-th cluster $\mathcal{C}_m$), with associated collective coordinates $\alpha_m$ and $f_m$, respectively, such that
\begin{equation} \label{eq:CC_ansatz}
 \bm{\phi}(t) \approx \sum_{m=1}^M \alpha_m(t) \bm{\Phi}^{(m)} + f_m(t) \bm{1}_{N_m},
 \end{equation}
 where typically $2M \ll N$.
 
 The method of collective coordinates \cite{Gottwald15, Gottwald17, HancockGottwald18} is in effect a Galerkin approximation, where the residual error made by the ansatz (\ref{eq:CC_ansatz}) is minimized and the minimization leads to a system of evolution equations for the collective coordinates $\alpha_m(t)$ and $f_m(t)$.
 
 The choice of basis functions is crucial. The shape mode $\bm{\Phi}^{(m)}$ can be chosen via linearization of the Kuramoto model (\ref{eq:full_KM}), restricted to oscillators in $\mathcal{C}_m$. For sufficiently large coupling strengths $K$, $\bm{\Phi}^{(m)}$ will solve the Kuramoto model to good accuracy (ignoring the interactions with any oscillators outside of $\mathcal{C}_m$). 
 
We follow the methods outlined previously \cite{Gottwald15, Gottwald17, HancockGottwald18} and derive a collective coordinate reduction for multimodal natural frequency distributions of the form (\ref{eq:multimodal_distribution}). We first present the reduction for a single synchronized cluster of oscillators, and then present results for several interacting clusters.

\subsection{Single cluster ansatz}

Linearizing the full Kuramoto model (\ref{eq:full_KM}) around $\phi_i - \phi_j = 0$ for all $i,j$ results in
\begin{equation} \label{eq:SCA_linearization}
\dot{\bm{\phi}} = \bm{\omega}  - \frac{K}{N} L \bm{\phi},
\end{equation}
where $L = D - A$ is the graph Laplacian, and $D$ is the diagonal degree matrix, i.e., $D_{ii}$ is the degree of node $i$. Note that $L$ has a nontrivial kernel with $L \bm{1}_N = 0$, associated with the invariance to a global constant phase shift. Global synchronization corresponds to all oscillators rotating at the mean natural frequency $\Omega = (1/N)\sum_i \omega_i$. Substituting $\dot{\bm{\phi}} = \Omega \bm{1}_N$ into (\ref{eq:SCA_linearization}) we obtain the global synchronization mode
\begin{equation} \label{eq:SCA_mode}
\hat{\bm{\phi}} = \frac{N}{K} L^+ \bm{\omega},
\end{equation}
where $L^+$ denotes the pseudoinverse of $L$, and we note that $L^+ \bm{1}_n = \bm{0}$. Therefore, the single cluster ansatz function is
\begin{equation} \label{eq:SCA_function}
\bm{\Phi} = \alpha(t) \hat{\bm{\phi}},
\end{equation}
with collective coordinate $\alpha(t)$. For all-to-all coupling, $L = N I_N - \bm{1}_N \bm{1}_N^T$. Therefore, $L^+ = \frac{1}{N}( I_N - \frac{1}{N} \bm{1}_N \bm{1}_N^T)$ \footnote{$L$ has a zero eigenvalue with multiplicity one, and a repeated eigenvalue $\lambda= N$ with multiplicity $N-1$. Diagonalizing $L$, we have $L=VDV^T$ where $D = \text{diag}(N,\dots,N,0)$ and so $L^+=VD^+V^T=\frac{1}{N^2}L$.} and 
\begin{equation} \nonumber
\hat{\bm{\phi}} = \frac{1}{K}(\bm{\omega} - \Omega \bm{1}_N).
\end{equation}
Note that for a single synchronized cluster, as a result of the phase shift invariance, we may assume, without loss of generality, that $\Omega = 0$. For multiple synchronized clusters, the different mean natural frequencies of each cluster must be accounted for, which we show in \S\ref{sec:MCA}.

The evolution equation for the collective coordinate $\alpha(t)$ can be found as a Galerkin approximation using the same approach as in previous studies. \cite{Gottwald15, Gottwald17, HancockGottwald18} The ansatz (\ref{eq:SCA_function}) is substituted into the Kuramoto model (\ref{eq:full_KM}), yielding a residual error
\begin{equation} \nonumber
\mathcal{E}_i = \dot{\alpha}\hat{\phi}_i - \omega_i - \frac{K}{N}\sum_{j=1}^N A_{ij} \sin\left(\alpha\left(\hat{\phi}_j - \hat{\phi}_i\right)\right).
\end{equation}
This residual error, which is a two-dimensional manifold parametrized by $\alpha$ and $\dot{\alpha}$, is minimized when it is orthogonal to the one-dimensional line $\alpha \hat{\bm{\phi}}$ that we are restricting the solution to. Setting $\bm{\mathcal{E}}\cdot \hat{\bm{\phi}} = 0$ we obtain an evolution equation for the collective coordinate $\alpha$
\begin{equation} \nonumber
\dot{\alpha} = \frac{\hat{\bm{\phi}}^T \bm{\omega}}{\hat{\bm{\phi}}^T \hat{\bm{\phi}}} + \frac{1}{\hat{\bm{\phi}}^T \hat{\bm{\phi}}} \frac{K}{N} \sum_{i,j=1}^N  \hat{\phi_i} A_{ij} \sin \left( \alpha\left(\hat{\phi}_j - \hat{\phi}_i\right)\right).
\end{equation}
For all-to-all coupling with mean frequency $\Omega = 0$, this simplifies to 
\begin{equation} \nonumber
\dot{\alpha} = K + \frac{K^2}{\Sigma^2 N^2} \sum_{i,j=1}^N  \omega_i \sin\left( \frac{\alpha}{K} \left( \omega_j - \omega_i \right) \right),
\end{equation}
where $\Sigma^2 = (1/N) \sum_i \omega_i^2$ is the variance of the natural frequencies. Setting $\beta = \alpha/K$, so that $\bm{\phi} \approx \beta \bm{\omega}$, yields
\begin{equation} \label{eq:SCA_evolution}
\dot{\beta} = 1 + \frac{K}{\Sigma^2 N^2} \sum_{i,j=1}^N \omega_i \sin\left( \beta \left( \omega_j - \omega_i \right) \right).
\end{equation}
Stationary points of (\ref{eq:SCA_evolution}) correspond to synchronized states for the Kuramoto model.

In the thermodynamic limit, $N\to \infty$, (\ref{eq:SCA_evolution}) becomes
\begin{equation} \label{eq:SCA_general_evolution_infinite}
\dot{\beta} = 1 + \frac{K}{\Sigma^2}  \iint \omega \sin \left( \beta \left( \eta - \omega \right) \right) g(\omega) g(\eta) d\omega d\eta = \mathcal{I}(\beta).
\end{equation}
For normally distributed natural frequencies, with mean zero and variance $\sigma^2$, we obtain
\begin{equation} \label{eq:SCA_evolution_infinite}
\mathcal{I}(\beta) = 1 - K \beta \exp \left( - \sigma^2 \beta^2 \right).
\end{equation}
Since $\mathcal{I}(0)=1$, it follows that $\beta$ has a stationary point if and only if $\mathcal{I}$ has a negative local minimum. Solving $\frac{d\mathcal{I}}{d\beta}=0$ and $\frac{d^2 \mathcal{I}}{d\beta^2}>0$ yields $\beta = (\sigma\sqrt{2})^{-1}$. Therefore, $\beta$ has a stationary point if and only if $\mathcal{I}\left(\left(\sigma\sqrt{2}\right)^{-1} \right) \leq 0$, which is equivalent to
\begin{equation} \label{eq:SCA_test_1}
K \geq \sigma \sqrt{2e}.
\end{equation}
If the condition (\ref{eq:SCA_test_1}) is satisfied, the oscillators synchronize and form a single cluster. 

The instantaneous order parameter for the collective coordinates can be calculated as 
\begin{equation} \label{eq:normal_order_param}
 r(t) = \exp\left(-\frac{\sigma^2 \beta^2}{2} \right).
\end{equation} 
This relation shows that $\beta$ measures the spread of the oscillators. Large values of $\beta$ correspond to small $r$, meaning the oscillators are evenly distributed on the circle, whereas small values of $\beta$ for which $|\bm{\Phi}| \ll 1$ correspond to $r \approx 1$, corresponding to tightly clustered oscillators.

For the multimodal natural frequency distribution (\ref{eq:multimodal_distribution}) with $M$ peaks we obtain
\begin{align} 
\mathcal{I}(\beta) = & 1 + \frac{K}{\Sigma^2} \sum_{i,j=1}^M  \gamma_i \gamma_j e^{-\frac{1}{2} \beta^2 (\sigma_i^2 + \sigma_j^2)} \times \nonumber \\
& \left[ \Omega_j \sin \left( \beta \left( \Omega_i - \Omega_j \right) \right) - \beta \sigma_j^2 \cos \left( \beta \left( \Omega_i - \Omega_j \right) \right) \right]. \nonumber
\end{align}
As for a unimodal distribution, $\mathcal{I}(0)=1$, and a stable stationary solution of (\ref{eq:SCA_general_evolution_infinite}), corresponding to global synchronization of oscillators, exists if and only if the minimum of $\mathcal{I}(\beta)$ (obtained numerically) is negative. Therefore, the condition for global synchronization is
\begin{equation} \label{eq:SCA_test_2}
\min_{\beta} \mathcal{I}(\beta) < 0.
\end{equation}

\subsection{Multiple cluster ansatz} \label{sec:MCA}

For multimodal frequency distributions, there is generally a range of $K$ values which are sufficiently large that oscillators form synchronized clusters, $\mathcal{C}_1,\dots,\mathcal{C}_M$, corresponding to each peak in the distribution, but which are not sufficiently large to allow for global synchronization. In such a case, we use a modified ansatz, which accounts for intracluster and intercluster dynamics. Note that while we are primarily concerned with clusters originating from a multimodal natural frequency distribution, the same analysis can be performed for topological clusters. \cite{HancockGottwald18} For oscillators in cluster $\mathcal{C}_m$, the intracluster dynamics is given by the restricted Kuramoto model
\begin{equation} \nonumber
\dot{\phi}_i^{(m)} = \omega_i^{(m)} + \frac{K}{N} \sum_{j\in \mathcal{C}_m} A_{ij} \sin\left(\phi_j^{(m)} - \phi_i^{(m)}\right),
\end{equation}
where for now we ignore the influence of oscillators belonging to different clusters $k \neq m$. Following the same linearization procedure as for the full Kuramoto model yields the intracluster mode
\begin{equation} \label{eq:cluster_mode}
\hat{\bm{\phi}}^{(m)} = \frac{N}{K}L_m^+ \bm{\omega}^{(m)},
\end{equation}
where $L_m^+$ is the pseudo-inverse of the graph Laplacian of the subgraph obtained by restricting to nodes in cluster $\mathcal{C}_m$. In the case of all-to-all coupling we obtain
\begin{equation}  \nonumber
\hat{\bm{\phi}}^{(m)} = \frac{N}{K N_m}\left( \bm{\omega}^{(m)} - \Omega_m \bm{1}_{N_m} \right),
\end{equation}
where $N_m$ is the number of oscillators in cluster $m$, and $\Omega_m$ is the mean frequency of cluster $m$. Note that for well separated peaks in the frequency distribution, such as the example in Fig.~\ref{fig:quadmodal_distro}, $N_m/N \approx \gamma_m$, where $\gamma_m$ is the weighting of peak $m$ in the natural frequency distribution (\ref{eq:multimodal_distribution}). 

The intracluster mode $\hat{\bm{\phi}}^{(m)}$ does not account for interactions with oscillators not belonging to cluster $m$. Therefore, $\hat{\bm{\phi}}^{(m)}$ does not capture the asymptotic dynamics of the system for large $K$, where the oscillators will globally synchronize and form a single cluster. For global synchronization, the single cluster ansatz $\hat{\bm{\phi}}$ in (\ref{eq:SCA_mode}) is a more appropriate mode. We remark that one can perform a Galerkin approximation valid for all coupling strengths by considering a linear superposition of the single cluster mode (\ref{eq:SCA_mode}), the superposition of all possible synchronized clusters (\ref{eq:cluster_mode}), as well as all possible mergings of synchronized clusters, each of these equipped with their own collective coordinate. However, since the advantage of employing the collective coordinate reduction is simplicity, which allows us to study the dynamics of the $N$-dimensional Kuramoto model, we prefer to use Galerkin approximations tailored for a particular dynamical range, parametrized by the coupling strength $K$.

When studying intercluster dynamics between cluster modes (\ref{eq:cluster_mode}), the Galerkin approximation needs to account for the mean phases of each cluster, denoted $f_m$. These phases vary in time due to interactions between clusters. Accounting for these phase interactions, and the possibility of all clusters merging into a single cluster, for oscillators in cluster $m$ we propose the ansatz
\begin{equation} \label{eq:MCA_full_ansatz}
\bm{\Phi}^{(m)} =\alpha \pi^{(m)} \hat{\bm{\phi}} + \alpha_m \hat{\bm{\phi}}^{(m)} + f_m \bm{1}_{N_m},
\end{equation}
where $\pi^{(m)}$ denotes projection onto the nodes in cluster $m$, i.e. $\pi^{(m)}(v_i)=v_i$ if $i\in \mathcal{C}_m$ and $\pi^{(m)}(v_i)=0$ if $i\notin \mathcal{C}_m$. Here $\alpha, \alpha_m$ and $f_m$, $m=1,\dots,M$, are the collective coordinates. As for the single cluster ansatz, the dynamics for the collective coordinates are obtained by substituting the ansatz (\ref{eq:MCA_full_ansatz}) into the Kuramoto model (\ref{eq:full_KM}) to determine the residual error. Then, to ensure errors are minimized, we require the error to be orthogonal to the restricted solution hyperplane, spanned by $\hat{\bm{\phi}}$, $\hat{\bm{\phi}}^{(m)}$ and $\bm{1}_{N_m}$. The condition that the residual error is orthogonal to $\hat{\bm{\phi}}$ is given by
\begin{equation} \label{eq:MCA_general_topology_1}
\hat{\bm{\phi}}^T \hat{\bm{\phi}} \dot{\alpha} + \sum_{m=1}^M (\pi^{(m)} \hat{\bm{\phi}})^T\left(  \hat{\bm{\phi}}^{(m)} \dot{\alpha}_m + \bm{1}_{N_m} \dot{f}_m \right) = \hat{\bm{\phi}}^T \bm{G}(\bm{\Phi}),
\end{equation}
where 
\begin{equation} \label{eq:MCA_general_topology_2}
\bm{G}(\bm{\Phi}) = \bm{\omega} + \left(\frac{K}{N}\sum_{j=1}^N \sin(\Phi_j-\Phi_i) \right)_{i=1,\dots,N}
\end{equation}
is the right hand side of the Kuramoto model (\ref{eq:full_KM}) in vector form. The condition that the residual error is orthogonal to $\hat{\bm{\phi}}^{(m)}$ is given by
\begin{equation} \label{eq:MCA_general_topology_3}
 (\hat{\bm{\phi}}^{(m)})^T \pi^{(m)} \hat{\bm{\phi}}  \dot{\alpha} +  (\hat{\bm{\phi}}^{(m)})^T  \hat{\bm{\phi}}^{(m)}  \dot{\alpha}_m =  (\hat{\bm{\phi}}^{(m)})^T \pi^{(m)}\bm{G}(\bm{\Phi}).
\end{equation}
(We note that since $\hat{\bm{\phi}}^{(m)}$ is orthogonal to $\bm{1}_{N_m}$ there is no $\dot{f}_m$ term). Lastly, the condition that the residual error is orthogonal to $\bm{1}_{N_m}$ is given by
\begin{equation} \nonumber
\bm{1}_{N_m}^T \pi^{(m)} \hat{\bm{\phi}}  \dot{\alpha} +  N_m  \dot{f}_m =  \bm{1}_{N_m}^T \pi^{(m)}\bm{G}(\bm{\Phi}).
\end{equation}
Equations (\ref{eq:MCA_general_topology_1})--(\ref{eq:MCA_general_topology_3}) form a system of linear equations
\begin{equation} \nonumber
\mathcal{A} \dot{\bm{x}} = \bm{b}(\bm{x}),
\end{equation}
where $\bm{x} = (\alpha, \alpha_1, \dots,\alpha_m,f_1,\dots,f_m)^T$ is the vector comprised of the collective coordinates. This linear system can be solved to find the evolution equations for each of the collective coordinates.

In the case of all-to-all coupling, the projection $\pi^{(m)}\hat{\bm{\phi}}$ (\ref{eq:SCA_mode}), the cluster modes $\hat{\bm{\phi}}^{(m)}$ (\ref{eq:cluster_mode}) and the constant vectors $\bm{1}_{N_m}$ are linearly dependent, and so the ansatz (\ref{eq:MCA_full_ansatz}) simplifies to
\begin{align} \label{eq:MCA_function}
\Phi^{(m)}_i &= \alpha \frac{\omega_i}{K}  + \alpha_m \left( \frac{N}{KN_m} \left( \omega_i - \Omega_m \right) \right) + f_m \nonumber \\	
	&= \beta_m \left(\omega_i - \Omega_m\right) + \tilde{f}_m,
\end{align}
where $\beta_m = \frac{1}{K} \left( \frac{N}{N_m} \alpha_m + \alpha\right)$ and $\tilde{f}_m = f_m + \alpha \frac{\Omega_m}{K}$. This means that the global synchronization ansatz (\ref{eq:SCA_mode}) can be fully described by the cluster modes (\ref{eq:cluster_mode}) with suitable mean phases of each mode, and so the collective coordinate $\alpha$ associated with global synchronization can effectively be ignored. \footnote{For Erd\H{o}s--Renyi networks with large number of oscillators, we observe that the projection $\pi^{(m)}\hat{\bm{\phi}}$ (\ref{eq:SCA_mode}), the cluster mode $\hat{\bm{\phi}}^{(m)}$ (\ref{eq:cluster_mode}) and the constant vector $\bm{1}_{N_m}$ are \emph{almost} linearly dependent, and so $\alpha$ can be ignored from the ansatz to leading order.} In essence, $\beta_m$ measures the spread of the oscillators within cluster $m$, and $\tilde{f}_m$ determines the collective phase of the cluster.

For the ansatz (\ref{eq:MCA_function}), the evolution equations for the collective coordinates obtained from (\ref{eq:MCA_general_topology_1})--(\ref{eq:MCA_general_topology_3}) become
\begin{widetext}
\begin{align}
\dot{\beta}_m &= 1 + \frac{1}{N_m\Sigma_m^2}\frac{K}{N} \sum_{k=1}^M \sum_{j\in \mathcal{C}_k} \sum_{i\in \mathcal{C}_m} \left( \omega^{(m)}_i - \Omega_m \right) \sin \left( \beta_k\left( \omega^{(k)}_j - \Omega_k \right) - \beta_m\left( \omega^{(m)}_i - \Omega_m \right) + f_k - f_m \right), \label{eq:MCA_dynamics_finite_1} \\ 
\dot{f}_m &= \Omega_m + \frac{1}{N_m}\frac{K}{N} \sum_{k=1}^M \sum_{j\in \mathcal{C}_k} \sum_{i\in \mathcal{C}_m}  \sin \left( \beta_k\left( \omega^{(k)}_j - \Omega_k \right) - \beta_m\left( \omega^{(m)}_i - \Omega_m \right) + f_k - f_m \right), \label{eq:MCA_dynamics_finite_2}
\end{align}
\end{widetext}
where we have dropped the tilde on $f_m$, and $\Sigma_m^2 = \frac{1}{N_m} \sum_{i\in \mathcal{C}_m} \left(\omega^{(m)}_i - \Omega_m\right)^2$ is the variance of the frequencies in cluster $m$. In the following, we consider all-to-all networks, unless stated otherwise, and consider (\ref{eq:MCA_dynamics_finite_1})--(\ref{eq:MCA_dynamics_finite_2}). Therefore, for $M$ peaks in the frequency distribution, there are $2M$ equations of motion. By introducing phase difference variables, $F_m = f_{m+1} - f_m$, we reduce the dimension of the system to $2M - 1$ degrees of freedom. This suggests that chaos may be possible as long as $M\geq 2$. However, as we will show, chaos is only possible if $M\geq 3$.

In the thermodynamic limit, $N\to \infty$, with a multimodal natural frequency distribution of the form (\ref{eq:multimodal_distribution}), $N_m/N \to \gamma_m$ (the weight of cluster $m$), and the evolution equations for the collective coordinates (\ref{eq:MCA_dynamics_finite_1})--(\ref{eq:MCA_dynamics_finite_2}) become
\begin{align}
\dot{\beta}_m &= 1 - K \beta_m e^{-\frac{\sigma_m^2 \beta_m^2}{2}} \sum_{k=1}^M \gamma_k e^{-\frac{\sigma_k^2 \beta_k^2}{2}} \cos\left( f_k - f_m \right), \label{eq:MCA_dynamics_infinite_1} \\
\dot{f}_m &= \Omega_m + K  e^{-\frac{\sigma_m^2 \beta_m^2}{2}} \sum_{k=1}^M \gamma_k e^{-\frac{\sigma_k^2 \beta_k^2}{2}} \sin\left( f_k - f_m \right). \label{eq:MCA_dynamics_infinite_2}
\end{align}
Note that for $M=1$, i.e., unimodal, normally distributed frequencies with $\gamma_1=1$ and $\Omega_1=0$, (\ref{eq:MCA_dynamics_infinite_1}) recovers the single cluster evolution equation (\ref{eq:SCA_evolution_infinite}), and (\ref{eq:MCA_dynamics_infinite_2}) is identically zero.

\subsection{Slow-fast splitting of the shape and phase coordinates} \label{sec:time_scale_splitting}

Each synchronized cluster viewed in isolation contains oscillators with normally distributed natural frequencies. Therefore, the instantaneous order parameter for each cluster is given in the thermodynamic limit (cf. (\ref{eq:normal_order_param})) by
\begin{equation} \nonumber
r_m(t) = \exp\left( - \frac{\sigma_m^2 \beta_m^2}{2} \right).
\end{equation}
Expressing the evolution equations (\ref{eq:MCA_dynamics_infinite_1})--(\ref{eq:MCA_dynamics_infinite_2}) for the collective coordinates $\beta_m$ and $f_m$ in terms of $r_m$, we obtain
\begin{align}
\dot{r}_m &= -\sigma_m r_m \sqrt{\log r_m^{-2}}\, \times \nonumber \\
 & \left(  1 - \frac{K r_m}{\sigma_m} \sqrt{\log r_m^{-2}}  \sum_{k=1}^M \gamma_k r_k \cos\left( f_k - f_m \right) \right), \label{eq:MCA_dynamics_infinite_3} \\
\dot{f}_m &= \Omega_m + K  r_m \sum_{k=1}^M \gamma_k r_k \sin\left( f_k - f_m \right). \label{eq:MCA_dynamics_infinite_4}
\end{align}

In the case that each cluster remains tightly clustered for all time, we have $r_m(t) = 1 - \epsilon_m(t)$, with $0<\epsilon_m(t) \ll 1$ for all $t$. This is ensured provided the $\sigma_m$ are sufficiently small, $K$ is sufficiently large (i.e., the condition (\ref{eq:SCA_test_1}) is satisfied for each $\sigma_m$), and the means $\Omega_m$ are sufficiently far apart relative to the coupling strength (i.e., condition (\ref{eq:SCA_test_2}) fails and global synchronization does not occur). Expanding (\ref{eq:MCA_dynamics_infinite_3})--(\ref{eq:MCA_dynamics_infinite_4}) in powers of $\epsilon$ yields
\begin{align}
\dot{r}_m \!&= \! -\epsilon_m^{1/2} \sigma_m \sqrt{2} + 2K \epsilon_m  \sum_{k=1}^M \gamma_k \cos\left( f_k \!- \!f_m \right)  + \mathcal{O}\!\left(\!\epsilon^{3/2}\right) \label{eq:MCA_dynamics_reduced_1} \\
\dot{f}_m &= \Omega_m + K  \sum_{k=1}^M \gamma_k \sin\left( f_k \!- \! f_m \right) + \mathcal{O}\left( \epsilon \right). \label{eq:MCA_dynamics_reduced_2}
\end{align}

We can view the order parameters $r_m$ as describing the intracluster dynamics and the phase coordinates $f_m$ as describing the intercluster dynamics. Since $\epsilon_m \ll 1$, the evolution equations (\ref{eq:MCA_dynamics_reduced_1})--(\ref{eq:MCA_dynamics_reduced_2}) for $r_m$ and $f_m$ reveal a time-scale splitting of the dynamics, whereby the order parameters $r_m$ evolve slowly, whereas the phase variables $f_m$ evolve on a fast time scale. The intercluster dynamics is, to first-order, decoupled from the intracluster dynamics (cf. (\ref{eq:MCA_dynamics_reduced_2})). Hence, the intercluster dynamics obeys a reduced, renormalized Kuramoto model. Since the reduced intercluster dynamics has $M-1$ degrees of freedom (taking into account a change to phase difference variables), chaos is only possible if $M \geq 4$. We label this type of chaos where clusters remain localized, with only small changes in their order parameter, as phase chaos. However, it is possible that one or more of the clusters intermittently break-up, such that $r_m \ll 1$ and $\epsilon_m$ is not small anymore. In such a case, there is significant interplay between the intracluster and intercluster dynamics. This will be discussed in more detail in \S\ref{sec:three_clusters}.

\section{Four clusters: collective phase chaos} \label{sec:four_clusters}

Phase chaos is typically observed in systems with multimodal natural frequency distributions with at least four peaks [cf. Fig.~\ref{fig:complex_order_param_trajectories}(b)]. The simplest case is to take four oscillators with natural frequencies equally spaced between $-1$ and $1$ and let them interact to produce chaotic dynamics. \cite{PopovychEtAl05, MaistrenkoEtAl05, MiritelloEtAl09, CarluEtAl18} One may then consider $N$ oscillators distributed over these four distinct natural frequencies, or, more generally, consider the natural frequency distribution of $M$ distinct mean frequencies $\Omega_m$
\begin{equation} \label{eq:dirac_delta_frequencies}
g(\omega) = \sum_{m=1}^M \frac{1}{M} \delta\left(\omega - \Omega_m\right),
\end{equation}
where $\delta(x)$ denotes the Dirac delta-function. If $N$ frequencies, $\omega_1,\dots,\omega_N$, are distributed equiprobably onto the $M$ mean frequencies $\Omega_m$, with $N$ divisible by $M$, then each mean frequency $\Omega_m$ is populated by $N/m$ oscillators with $\omega_i = \Omega_m$. That is, we can relabel such that $\omega_1,\dots,\omega_{N/M} = \Omega_1$, $\omega_{N/M+1},\dots,\omega_{2N/M} = \Omega_2$, and so on. The Kuramoto model (\ref{eq:full_KM}) for oscillators with natural frequency $\Omega_m$ and all-to-all coupling in this case becomes
\begin{equation} \label{eq:dirac_delta_KM}
\dot{\phi}^{(m)}_i = \Omega_m + \frac{K}{N} \sum_{k=1}^M \sum_{j=1}^{N/M} \sin\left( \phi^{(k)}_j - \phi^{(m)}_i \right).
\end{equation}
Since the coupling is all-to-all, oscillators with the same natural frequency will synchronize, such that $\bm{\phi}^{(m)}(t) = f_m(t) \bm{1}_{N/M}$ and (\ref{eq:dirac_delta_KM}) becomes
\begin{equation} \label{eq:dirac_delta_reduced_KM}
\dot{f}_m = \Omega_m + \frac{K}{M} \sum_{k=1}^M \sin\left( f_k - f_m \right), 
\end{equation}
which is of the exact form as the Kuramoto model for $M$ oscillators. Hence, chaos is expected for arbitrarily many oscillators if their natural frequencies are distributed according to (\ref{eq:dirac_delta_frequencies}) with $M\geq 4$ with equally spaced $\Omega_m$. Note that the evolution equation for the phases $f_m$ (\ref{eq:dirac_delta_reduced_KM}) is equivalent to the collective coordinate equations for $M$ clusters (\ref{eq:MCA_dynamics_reduced_1})--(\ref{eq:MCA_dynamics_reduced_2}) in the limit $\epsilon_m \to 0$, which is the limit of perfectly synchronized clusters, with identical phases within each cluster.
\\

Considering the Dirac $\delta$-function as the limit of normal distributions, i.e., $\delta(x) = \lim_{\sigma \to 0} \mathcal{N}(0,\sigma)$, we expect multimodal distributions of the form (\ref{eq:multimodal_distribution}) with $M\geq 4$ to yield phase chaos for sufficiently small $\sigma$ and sufficiently large spacings between peaks in the natural frequency distribution, $|\Omega_{m+1}-\Omega_m|$. Our focus in this section is to explore the collective dynamics of the Kuramoto model for natural frequency distributions $g(\omega)$ of the form (\ref{eq:multimodal_distribution}) with identical weights $\gamma_m=1/4$, identical standard deviations $\sigma_m=\sigma$, and equally spaced means $\Omega_m = -1 + 2(m-1)/3$ for $m=1,\dots,4$, that is
\begin{equation} \nonumber
g(\omega) =  \frac{1}{4\sigma}  \sum_{m=1}^4P\left(\frac{\omega-\Omega_m}{\sigma}\right),
\end{equation}
where $P$ is the standard normal distribution.

\begin{figure*}[tbp]
\centering
\includegraphics[width=0.65\textwidth]{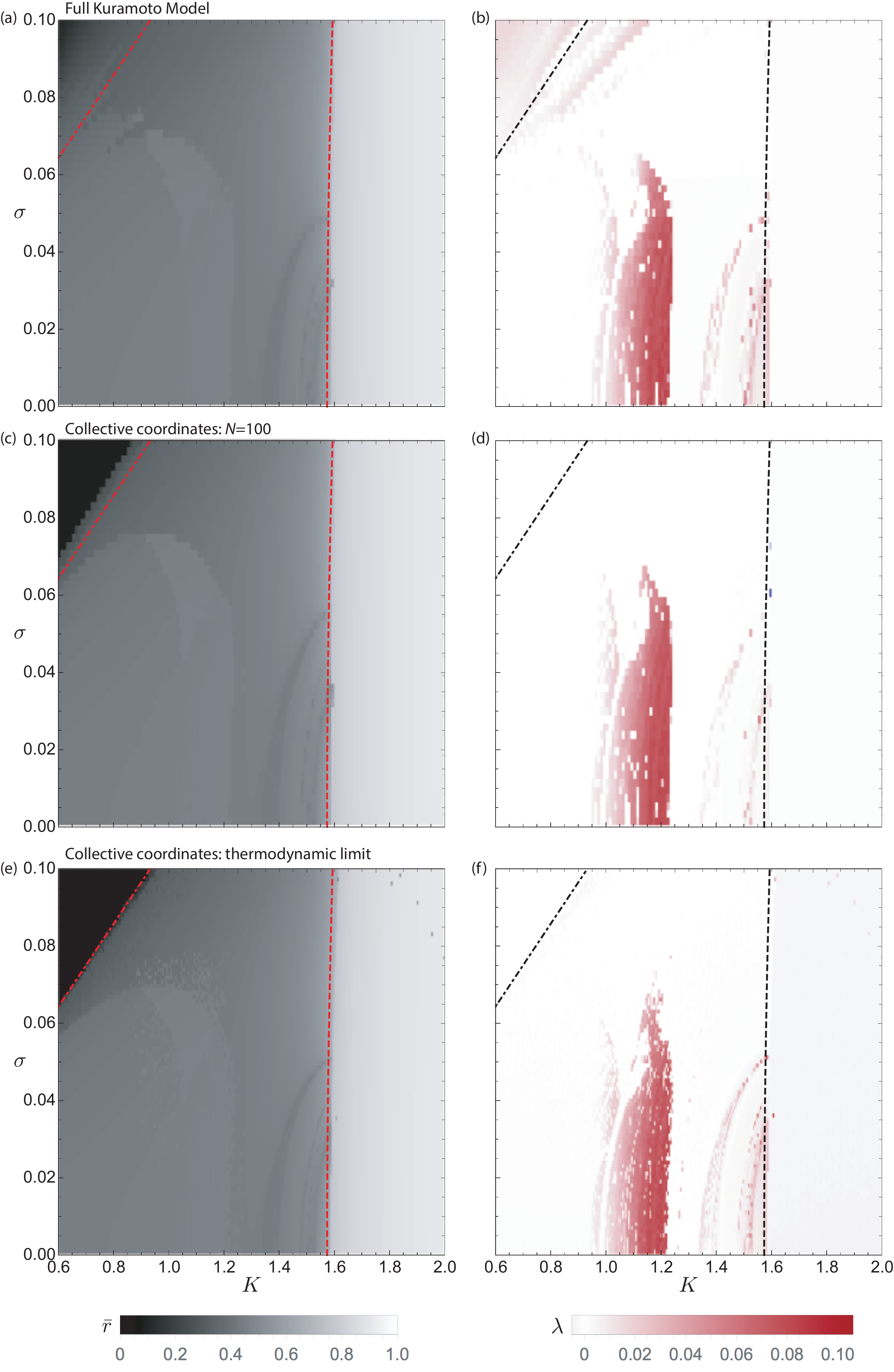}
\caption{Time averaged order parameter, $\bar{r}$ (left column), and leading Lyapunov exponent, $\lambda_1$ (right column), for a range of coupling strengths $K$ and multimodal natural frequency distributions with four peaks and means $\Omega_m = -1 + 2(m-1)/3$, weights $\gamma_m=1/4$, and identical standard deviations $\sigma_m=\sigma$, for $m=1,\dots,4$. (a),(b)~Full Kuramoto model (\ref{eq:full_KM}) with $N=100$ oscillators. (c),(d)~Collective coordinate model (\ref{eq:MCA_dynamics_finite_1})--(\ref{eq:MCA_dynamics_finite_2}) with $M=4$ and $N=100$. (e),(f)~Collective coordinate model in the thermodynamic limit (\ref{eq:MCA_dynamics_infinite_1})--(\ref{eq:MCA_dynamics_infinite_2}) with $M=4$. The dotted-dashed lines denote the condition (\ref{eq:SCA_test_1}), $K=\sigma \sqrt{2e}$, for synchronized clusters. The dashed, approximately vertical, curves denote the condition (\ref{eq:SCA_test_2}) for global synchronization.}
\label{fig:four_cluster_rbar+lyapunov}
\end{figure*}

We now numerically explore these cases for the full Kuramoto model (\ref{eq:full_KM}) with $N=100$ oscillators; and shall compare our results with the reduced collective coordinate description (\ref{eq:MCA_dynamics_finite_1})--(\ref{eq:MCA_dynamics_finite_2}) for $N=100$ oscillators as well as with the reduced collective coordinate description (\ref{eq:MCA_dynamics_infinite_1})--(\ref{eq:MCA_dynamics_infinite_2}) in the thermodynamic limit of infinitely many oscillators. The collective coordinate systems involve 7 degrees of freedom: four shape parameters $\beta_m$ and three phase-difference variables $f_{m+1}- f_m$ (the evolution equations, however, are written for $f_m$ and hence are 8-dimensional). We shall see that the collective coordinate equations provide a reduced model that allows for a quantitative description of the chaotic dynamics of the Kuramoto model, and, in particular, for the estimation of the Lyapunov exponents of the full Kuramoto model. We compute and compare the time-averaged order parameter $\bar{r}$ and the leading Lyapunov exponent $\lambda$ across a multitude of cases for different coupling strengths $K$ and for different standard deviations of the natural frequency distribution $\sigma$.

Before discussing the results on the leading Lyapunov exponent we shall describe the dependence of $\bar{r}$ on $K$ and $\sigma$, shown in the left column of Fig.~\ref{fig:four_cluster_rbar+lyapunov}. Shown is the order parameter $\bar{r}$ for the full Kuramoto model (\ref{eq:full_KM}) with $N=100$ oscillators [Fig.~\ref{fig:four_cluster_rbar+lyapunov}(a)], the 8D collective coordinate model (\ref{eq:MCA_dynamics_finite_1})--(\ref{eq:MCA_dynamics_finite_2}) with $N=100$ oscillators [Fig.~\ref{fig:four_cluster_rbar+lyapunov}(c)], and the 8D collective coordinate model with infinitely many oscillators (\ref{eq:MCA_dynamics_infinite_1})--(\ref{eq:MCA_dynamics_infinite_2}) [Fig.~\ref{fig:four_cluster_rbar+lyapunov}(e)]. We see good quantitative agreement between all three models throughout most of the parameter space. All three models show transitions from $\bar{r}\approx 0.45$ to $\bar{r} \approx 1$ near $K\approx 1.58$, which is the transition from four synchronized clusters to global synchronization with one synchronized cluster. This transition can be predicted by the collective coordinate ansatz, using the single cluster ansatz (\ref{eq:SCA_function}) applied to the full distribution $g(\omega)$. The transition curve is given by the condition (\ref{eq:SCA_test_2}) for global synchronization, and is shown by the dashed, approximately vertical, curves in Fig.~\ref{fig:four_cluster_rbar+lyapunov}. The transition from the incoherent state ($\bar{r}\approx 0$) to the synchronized cluster state ($\bar{r} \approx 0.45$) is predicted by the line $K = \sigma \sqrt{2 e} $ (dot-dashed in Fig.~\ref{fig:four_cluster_rbar+lyapunov}), which derives from condition (\ref{eq:SCA_test_1}) for the collective coordinate ansatz. However, this line does not accurately capture the transition from incoherence to synchronized clusters in the full Kuramoto model with $N=100$ oscillators [cf. Fig.~\ref{fig:four_cluster_rbar+lyapunov}(a)] for which the transition occurs at lower values of $K$. This discrepancy is due to the fact that the collective coordinate models (\ref{eq:MCA_dynamics_finite_1})--(\ref{eq:MCA_dynamics_finite_2}) and (\ref{eq:MCA_dynamics_infinite_1})--(\ref{eq:MCA_dynamics_infinite_2}) do not account for partial synchronization of the clusters. In the full Kuramoto model (\ref{eq:full_KM}) the transition from the incoherent state to a partially synchronized state is a soft second-order phase transition whereby, upon increasing the coupling strength, more and more oscillators with natural frequencies close to the mean frequency mutually synchronize until at a critical coupling strength all oscillators in a cluster have synchronized. Although this can be quantitatively described by the collective coordinate ansatz \cite{Gottwald15, HancockGottwald18} we knowingly do not account for this in our simulations here to limit the computational cost of the parametric sweep.

It is remarkable that the collective coordinate models   --- (\ref{eq:MCA_dynamics_finite_1})--(\ref{eq:MCA_dynamics_finite_2}) for $N=100$ oscillators [Fig.~\ref{fig:four_cluster_rbar+lyapunov}(d)] and (\ref{eq:MCA_dynamics_infinite_1})--(\ref{eq:MCA_dynamics_infinite_2}) for $N\to \infty$ [Fig.~\ref{fig:four_cluster_rbar+lyapunov}(f)] --- reproduce the leading Lyapunov exponent $\lambda$ of the full Kuramoto model (\ref{eq:full_KM}) [Fig.~\ref{fig:four_cluster_rbar+lyapunov}(b)] with good quantitative agreement. In particular, there is a chaotic ``bubble'' within the region with four synchronized clusters (between the dot-dashed and dashed curves) whose width shrinks as $\sigma$ increases. The occurrence of partial synchronization of clusters in the full Kuramoto model with $N=100$ results in a positive Lyapunov exponent above and near to the dot-dashed line in Fig.~\ref{fig:four_cluster_rbar+lyapunov}(b), which is not captured by the collective coordinate models [Fig.~\ref{fig:four_cluster_rbar+lyapunov}(d) and Fig.~\ref{fig:four_cluster_rbar+lyapunov}(f)]. This difference is due to complex interactions between the synchronized clusters and the small number of oscillators that do not synchronize, which are not accounted for by the collective coordinate models.

\begin{figure}[tbp]
\centering
\includegraphics[width=\columnwidth]{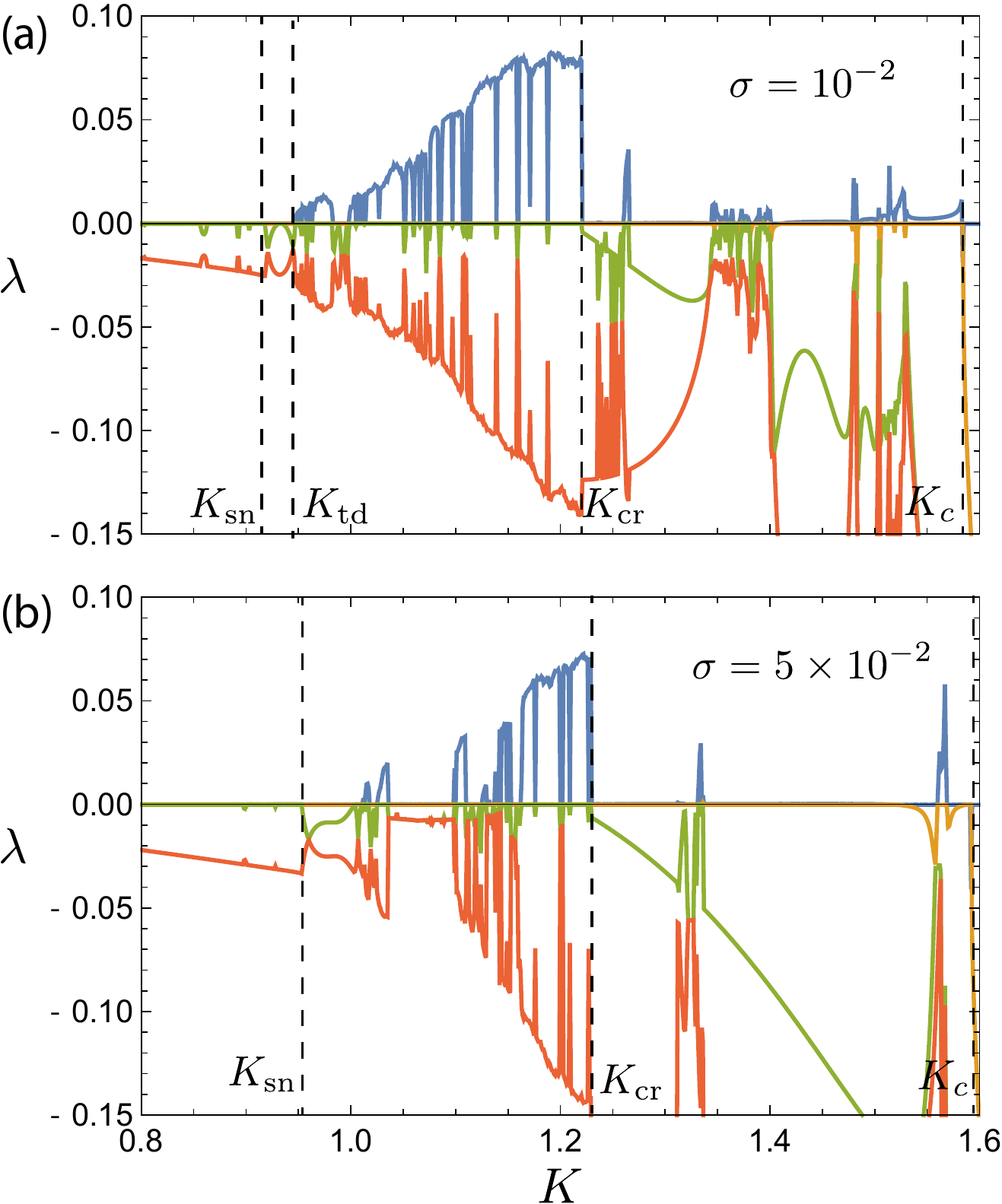}
\caption{The first four Lyapunov exponents of the collective coordinate model (\ref{eq:MCA_dynamics_infinite_1})--(\ref{eq:MCA_dynamics_infinite_2}) for a range of coupling strengths $K$ and multimodal natural frequency distributions with four peaks and means $\Omega_m = -1 + 2(m-1)/3$, weights $\gamma_m=1/4$, and identical standard deviations $\sigma_m=\sigma$, for $m=1,\dots,4$, (a)~$\sigma=10^{-2}$, (b)~$\sigma=5\times 10^{-2}$. }
\label{fig:four_cluster_lyapunov_exponents}
\end{figure}

In the limit as $\sigma \to 0$, the dynamics of four interacting clusters becomes equivalent to the dynamics of four interacting oscillators (cf. (\ref{eq:dirac_delta_reduced_KM})), which has been studied extensively by Maistrenko \emph{et al.} \cite{MaistrenkoEtAl05} and Popovych \emph{et al.} \cite{PopovychEtAl05} Following the approach of previous studies, we consider the first four Lyapunov exponents of the collective coordinate model (\ref{eq:MCA_dynamics_infinite_1})--(\ref{eq:MCA_dynamics_infinite_2}). For small values of $\sigma < 10^{-2}$ we obtain Lyapunov exponents that are qualitatively the same as those observed for four individual oscillators (compare Fig.~\ref{fig:four_cluster_lyapunov_exponents}(a) with Fig.~1(a) in Ref.~\citen{PopovychEtAl05}). Therefore, for these small values of $\sigma$ the bifurcation sequence is essentially the same as for four individual oscillators. At $K=K_\text{sn}\approx 0.91$ there is a saddle-node bifurcation, which transitions from quasiperiodic to periodic dynamics. At $K=K_\text{td}\approx 0.94$ there is a transition to chaos via the Afraimovich-Shilnikov torus destruction scenario. \cite{AfraimovichShilnikov91} At $K=K_\text{cr}\approx 1.22$ the chaotic attractor is destroyed in a boundary crisis, yielding a chaotic saddle. Lastly, at $K=K_c \approx 1.58$ the transition to global synchronization occurs. There are many periodic regions observed within the chaotic region $K_\text{td}<K<K_\text{cr}$, and also near $K=1.5$, which correspond to the resonances discussed by Maistrenko \emph{et al.} \cite{MaistrenkoEtAl05} The resonances within the chaotic region can also be observed within the chaotic bubble shown in the right plots of Fig.~\ref{fig:four_cluster_rbar+lyapunov}, evident as white bands ($\lambda_1 = 0$) that extend approximately vertically from the horizontal axis $\sigma = 0$ (most clearly seen in Fig.~\ref{fig:four_cluster_rbar+lyapunov}(f) which has the highest resolution). The resonances near $K=1.5$ can be seen in the right plots of Fig.~\ref{fig:four_cluster_rbar+lyapunov} as thin bands of positive largest Lyapunov exponent. For larger values of $\sigma$, such as $\sigma = 5\times 10^{-2}$ shown in Fig.~\ref{fig:four_cluster_lyapunov_exponents}(b), we see similar dynamics, but there are some key differences. First, the chaotic window is smaller, and is punctuated by a large periodic region near $K=1.05$. In addition, there appears to be only one resonance near $K=1.5$.

The complex bifurcation structure shown in Fig.~\ref{fig:four_cluster_lyapunov_exponents} also explains the discontinuous transition curves between different shades of gray in the plots for $\bar{r}$ (left plots of Fig.~\ref{fig:four_cluster_rbar+lyapunov}). These transitions are due to bifurcations between different stable chaotic, periodic, and quasiperiodic states.

\section{Three clusters: chaos via intermittent cluster desynchronization} \label{sec:three_clusters}

\begin{figure}[tbp]
\centering
\includegraphics[width=\columnwidth]{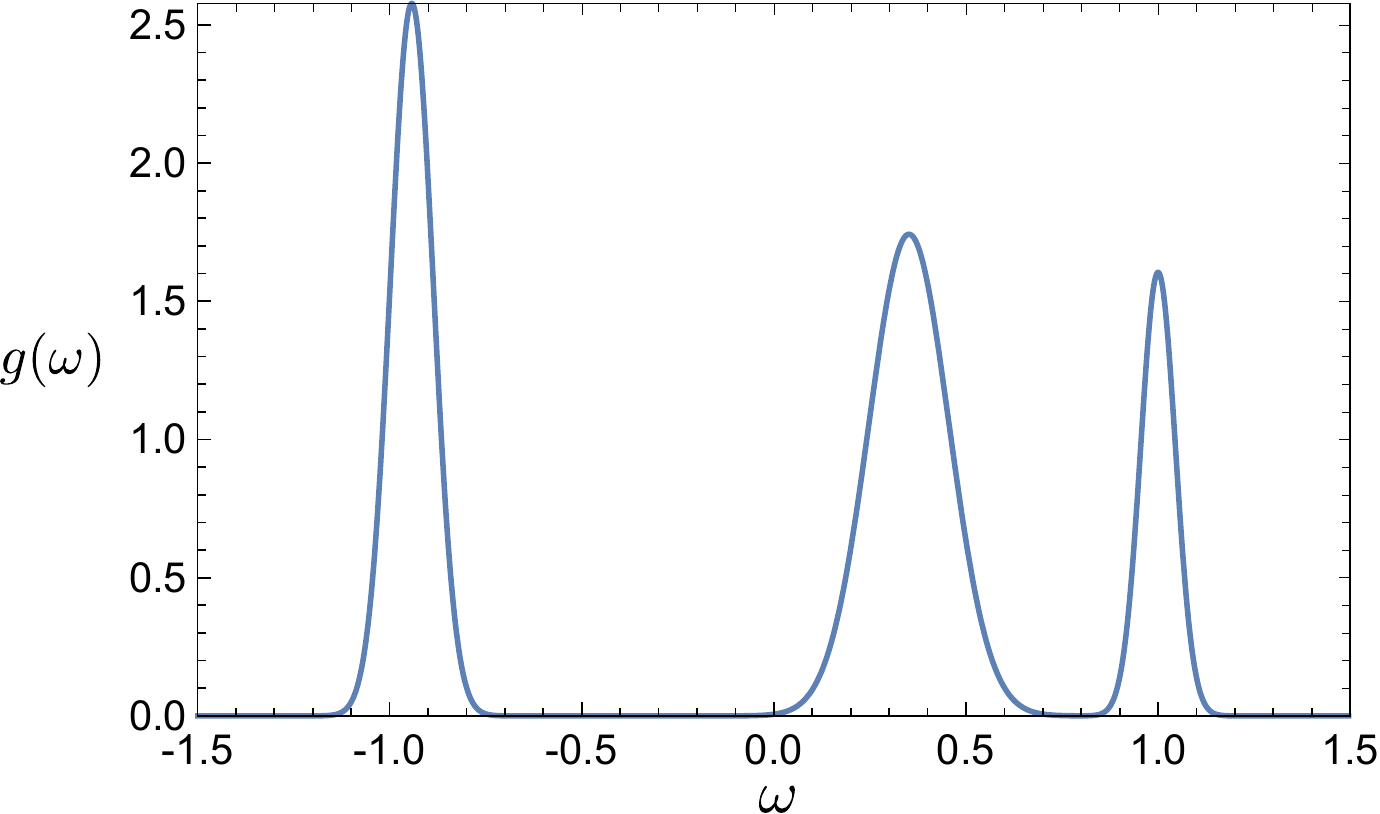}
\caption{Trimodal natural frequency distribution that results in chaotic dynamics of the Kuramoto model (\ref{eq:full_KM}). Parameters in the distribution function (\ref{eq:multimodal_distribution}) are chosen as $(\sigma_1,\sigma_2,\sigma_3) = (0.05617, 0.1042, 0.04521 )$, $(\Omega_1, \Omega_2, \Omega_3) = ( -0.9423, 0.3517, 1)$, and $(\gamma_1, \gamma_2, \gamma_3) = ( 0.3628, 0.4552, 0.1818 )$.}
\label{fig:trimodal_PDF}
\end{figure}

\begin{figure*}[tbp]
\centering
\includegraphics[width=\textwidth]{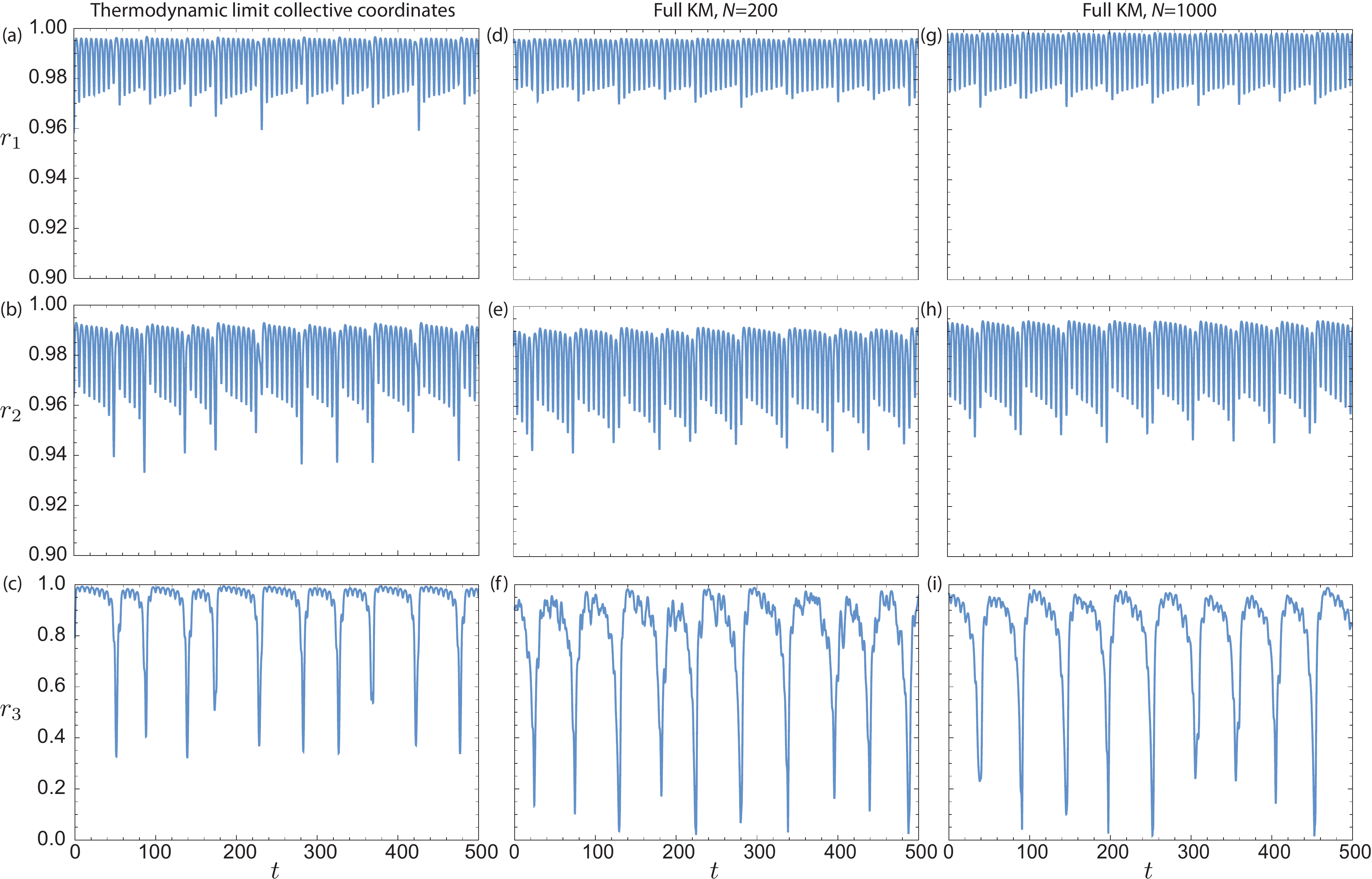}
\caption{Time series of the cluster order parameters $r_1, r_2$ and $r_3$ for the trimodal natural frequency distribution shown in Fig.~\ref{fig:trimodal_PDF}. Note the different scales on the vertical axis for $r_3$ compared to $r_1$ and $r_2$. (a--c)~Collective coordinate reduction in the thermodynamic limit (\ref{eq:MCA_dynamics_infinite_3})--(\ref{eq:MCA_dynamics_infinite_4}) with $M=3$. (d--f)~Full Kuramoto model (\ref{eq:full_KM}) with $N=200$ oscillators drawn equiprobably from the distribution shown in Fig.~\ref{fig:trimodal_PDF}. (g--i)~Full Kuramoto model (\ref{eq:full_KM}) with $N=1000$ oscillators drawn equiprobably from the distribution shown in Fig.~\ref{fig:trimodal_PDF}.}
\label{fig:trimodal_TS}
\end{figure*}

\begin{figure*}[tbp]
\centering
\includegraphics[width=\textwidth]{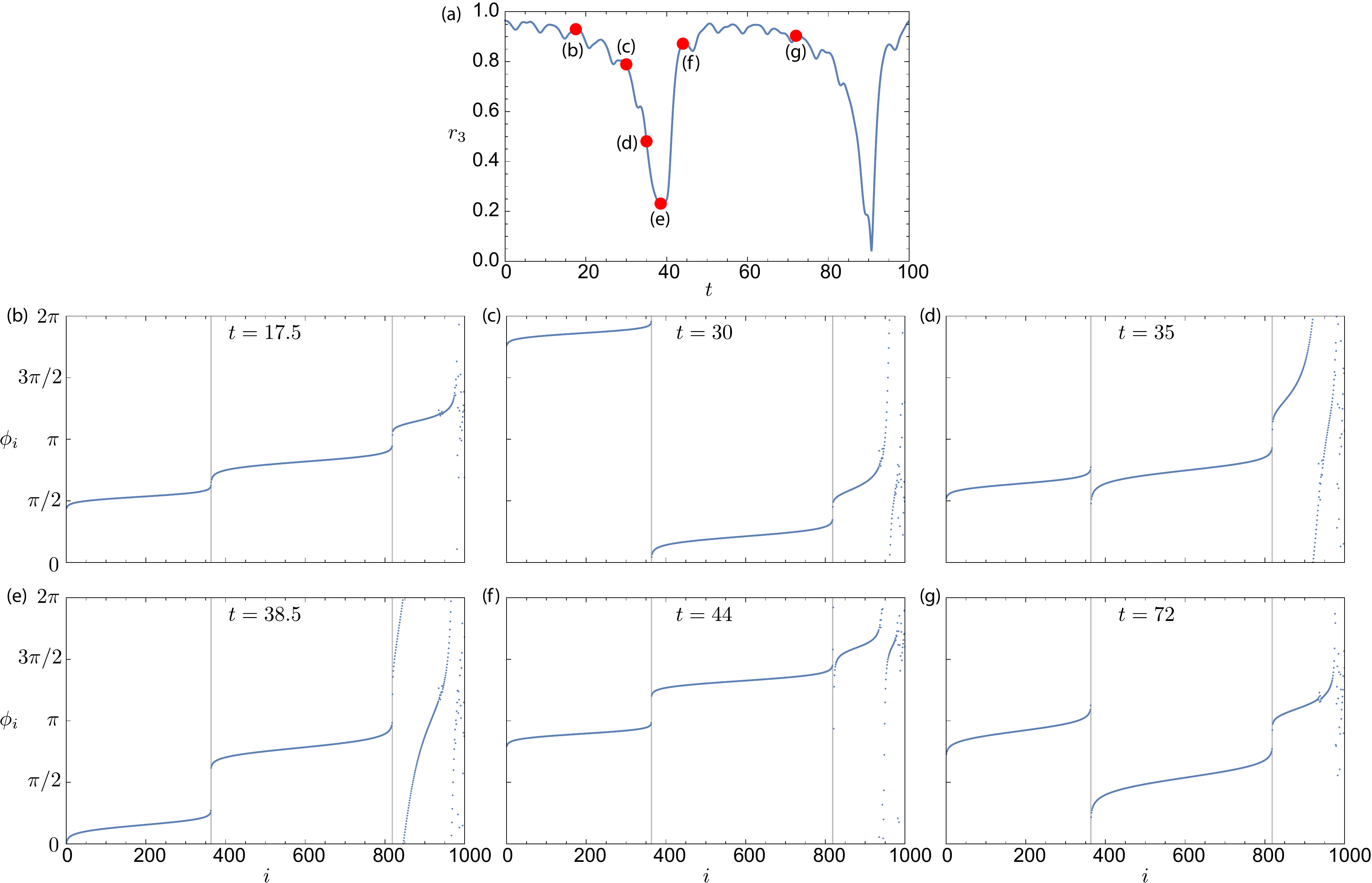}
\caption{(a)~Time series of $r_3$ for the full Kuramoto model (\ref{eq:full_KM}) with $N=1000$ oscillators [the same as Fig.~\ref{fig:trimodal_TS}(i)]. The labels of the red circles correspond to the snapshots of the oscillator phases $\phi_i$, shown in (b--g) illustrating the intermittent desynchronization of the third cluster from a coherent cluster that is phase-locked with the second cluster (b), to a desynchronized state (c)--(f), and back to a synchronized state (g). (Multimedia view)}
\label{fig:trimodal_snapshots}
\end{figure*}

For three clusters, as discussed previously, if the reduction and time-scale splitting shown in (\ref{eq:MCA_dynamics_reduced_1})--(\ref{eq:MCA_dynamics_reduced_2}) is valid, the dynamics is essentially phase dynamics of three oscillators, excluding chaotic dynamics because there are only two degrees of freedom (recall that due to the phase-gauge invariance of the Kuramoto model, we may assume without loss of generality that $\sum_i f_i = 0$). However, the time-scale splitting requires $\epsilon_m = 1 - r_m \ll 1$ for \emph{all} time. If this is not true, e.g. one cluster intermittently desynchronizes, then chaos is possible.

As an example, consider the trimodal natural frequency distribution shown in Fig.~\ref{fig:trimodal_PDF}. Simulating the 6D collective coordinate model, (\ref{eq:MCA_dynamics_infinite_3})--(\ref{eq:MCA_dynamics_infinite_4}), for $K= 1.205$ we find a positive largest Lyapunov exponent, $\lambda = 0.036$, as well as time-averaged cluster-wise order parameters $\bar{r}_1 = 0.989$, $\bar{r}_2 = 0.981$, $\bar{r}_3 = 0.918$. Hence the system is both chaotic and collectively organized. While $\bar{r}_3$ is close to one, and, hence, the cluster would be considered synchronized, the time series for $r_3(t)$, shown in Fig.~\ref{fig:trimodal_TS}(c), intermittently dips to values around $0.5$, showing that the cluster intermittently desynchronizes, with oscillators spreading over the entire circle. Therefore, we cannot say that $\epsilon_3$ is close to zero for all time, meaning the time-scale splitting is invalid for $r_3$, and, hence, chaos is possible.

This intermittent desynchronization phenomenon predicted by our collective coordinate reduction is confirmed in the full Kuramoto model (\ref{eq:full_KM}). For $N=200$ oscillators, with natural frequencies drawn equiprobably from the distribution $g(\omega)$ shown in Fig.~\ref{fig:trimodal_PDF}, we compute the leading Lyapunov exponent as  $\lambda = 0.039$, which is within $10\%$ of the Lyapunov exponent computed using collective coordinates in the thermodynamic limit, (\ref{eq:MCA_dynamics_infinite_3})--(\ref{eq:MCA_dynamics_infinite_4}), which has $\lambda = 0.036$. Furthermore, the time-series of $r_1, r_2, r_3$, shown in Fig.~\ref{fig:trimodal_TS}(d--f), are qualitatively similar to those shown in Fig.~\ref{fig:trimodal_TS}(a--c). In particular, $r_1$ and $r_2$ remain close to $1$ for all time, whereas $r_3$ experiences intermittent dips. The dips occur in the collective coordinate model (\ref{eq:MCA_dynamics_infinite_3})--(\ref{eq:MCA_dynamics_infinite_4}) with an average period of $48.9$, compared to an average period of $52.6$ in the full Kuramoto model. For the full Kuramoto model (\ref{eq:full_KM}) with $N=1000$ oscillators, which is closer to the thermodynamic limit, the dips occur at the same frequency as with $N=200$, i.e., with a period of $52.6$, and the time series of $r_{1,2,3}$, shown in Fig.~\ref{fig:trimodal_TS}(g--i), are even more similar to the collective coordinate model in the thermodynamic limit (\ref{eq:MCA_dynamics_infinite_3})--(\ref{eq:MCA_dynamics_infinite_4}), shown in Fig.~\ref{fig:trimodal_TS}(a--c), in that the dynamics between the dips becomes more regular, with high frequency oscillations and a slow negative trend. The collective coordinate model is representative of the full Kuramoto model, and has the advantage of being more analytically tractable.
\\

We now investigate more closely the nature of this type of chaotic dynamics and how it is generated. We first describe qualitatively the dynamics of a single desynchronization event in the full Kuramoto model (\ref{eq:full_KM}). We then show that these desynchronization events can be resolved by considering further reductions of the collective coordinate equations (\ref{eq:MCA_dynamics_infinite_3})--(\ref{eq:MCA_dynamics_infinite_4}). This collective coordinate reduction is then used to show that chaos via intermittent desynchronization is a robust phenomenon.

We describe the dynamics of a desynchronization event qualitatively using the snapshots of the phases of oscillators shown in Fig.~\ref{fig:trimodal_snapshots}(b--g) (Multimedia view), which correspond to the red points marked on the time series of $r_3$ shown in Fig.~\ref{fig:trimodal_snapshots}(a). In the lead-up to a dip in $r_3$, the second and third clusters are phase-locked, with an approximately constant phase difference $F_2 = f_3 - f_2$. However, each time the first cluster passes by the second cluster, the second cluster slows down, which causes a small increase in the phase separation between the second and the third clusters, implying a small increase in $F_2$, as shown in Fig.~\ref{fig:trimodal_snapshots}(b) and Fig.~\ref{fig:trimodal_snapshots}(c) as a small increase in separation between the second and third clusters. Eventually, a critical point is reached, such that the oscillators in the third cluster that are furthest from the second cluster [those with the highest natural frequencies, closest to $i=1000$ in Fig.~\ref{fig:trimodal_snapshots}(b--i)] begin to desynchronize with the rest of the oscillators in the cluster, as shown in the transition from Fig.~\ref{fig:trimodal_snapshots}(b) to Fig.~\ref{fig:trimodal_snapshots}(d). This desynchronization results in the oscillators in the third cluster wrapping around and covering the entire circle, and corresponds to a sharp dip in $r_3$, as shown in Fig.~\ref{fig:trimodal_snapshots}(a). The desynchronization of the third cluster occurs as a traveling front, starting first with the oscillators with highest natural frequency, traveling down to the oscillators with the lowest frequency. The oscillators in the third cluster eventually cover the entire circle, and those with the highest natural frequencies (furthest to the right in Fig.~\ref{fig:trimodal_snapshots}) overtake those with the lowest natural frequencies (furthest to the left in Fig.~\ref{fig:trimodal_snapshots}), meaning they experience additional revolutions during each ``dip'' event. Once the oscillators in the third cluster with lowest natural frequencies catch up with the second cluster, the third cluster re-synchronizes, as shown in Fig.~\ref{fig:trimodal_snapshots}(f) and Fig.~\ref{fig:trimodal_snapshots}(g), once again becoming phase-locked with the second cluster, and the process repeats.

\begin{figure}[tbp]
\centering
\includegraphics[width=\columnwidth]{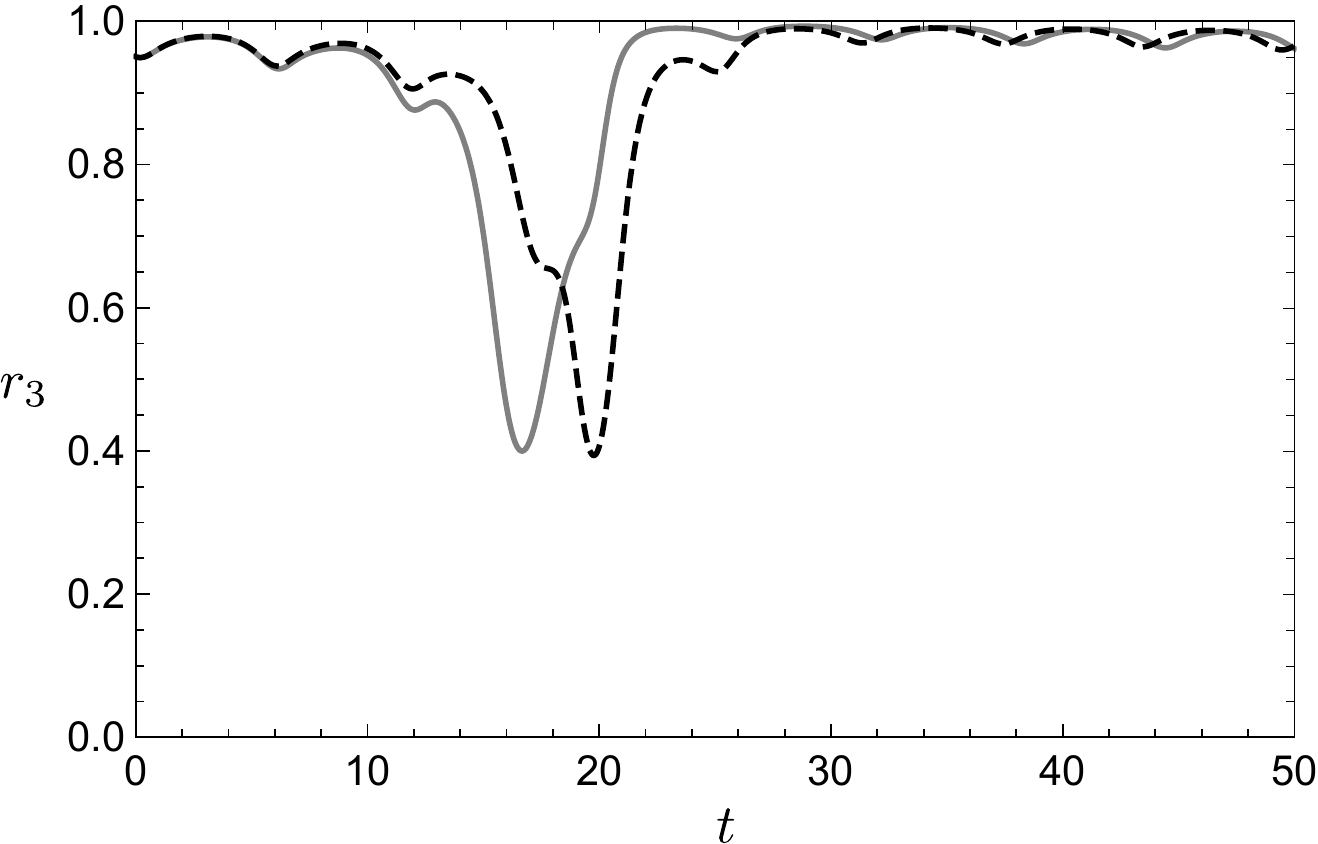}
\caption{Time series of $r_3$ for the 6D collective coordinates three-cluster equations (\ref{eq:MCA_dynamics_infinite_3})--(\ref{eq:MCA_dynamics_infinite_4}) with $M=3$ (solid gray), and the reduced 3D collective coordinate system (\ref{eq:trimodal_dynamics_3D_1})--(\ref{eq:trimodal_dynamics_3D_3}) (dashed black).}
\label{fig:trimodal_5D-vs-3D_time_series}
\end{figure}

\begin{figure}[tbp]
\centering
\includegraphics[width=\columnwidth]{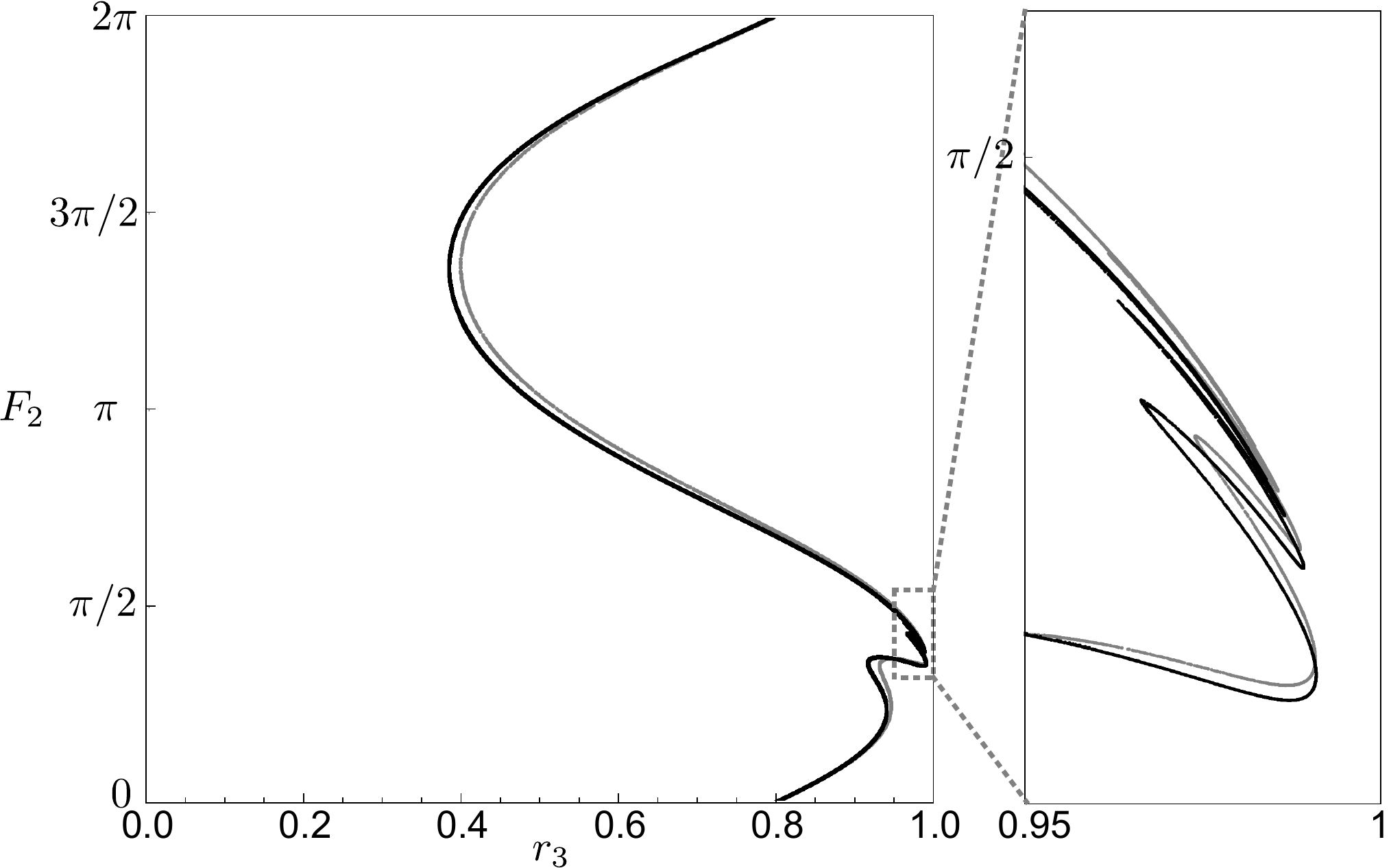}
\caption{Poincar\'{e} section of the collective coordinate dynamics for the trimodal natural frequency distribution shown in Fig.~\ref{fig:trimodal_PDF} through the plane $F_1=0$, shown in the $(r_3, F_2)$-plane. Shown are results for the 6D collective coordinate equations (\ref{eq:MCA_dynamics_infinite_3})--(\ref{eq:MCA_dynamics_infinite_4}) with $M=3$ (gray), and for the reduced 3D system (\ref{eq:trimodal_dynamics_3D_1})--(\ref{eq:trimodal_dynamics_3D_3} (black). The zoomed in region shows a fractal folding pattern for both models, indicating the presence of chaos.}
\label{fig:trimodal_psection}
\end{figure}

We now use collective coordinate reductions to analyze the dynamical scenario described above. As seen in Fig.~\ref{fig:trimodal_TS}, $r_1(t)$ and $r_2(t)$ are close to one for all time, demonstrating that the time-scale splitting remains valid for those variables. This suggests that we may set $r_1(t) = \bar{r}_1$ and $r_2(t) = \bar{r}_2$ as constant in the collective coordinate equations (\ref{eq:MCA_dynamics_infinite_3})--(\ref{eq:MCA_dynamics_infinite_4}). Then the collective coordinate model reduces to a system of four fast variables, $r_3, f_{1,2,3}$, with three degrees of freedom (again since, without loss of generality, $\sum_i f_i = 0$). The evolution equations become
\begin{widetext}
\begin{align}
\dot{r}_3 &= -\sigma_3 r_3 \sqrt{-2\log r_3} \left(  1 - \frac{K r_3}{\sigma_3} \sqrt{-2\log r_3} \left(\gamma_1 \bar{r}_1 \cos\left(F_1 + F_2\right) + \gamma_2 \bar{r}_2 \cos(F_2) + \gamma_3 r_3 \right) \right) \label{eq:trimodal_dynamics_3D_1} \\
\dot{F}_1 &= \Delta\Omega_1 + K \left[ -(1-\gamma_3) \bar{r}_1 \bar{r}_2 \sin F_1 + \gamma_3 \bar{r}_2 r_3 \sin F_2 - \gamma_3 \bar{r}_1 r_3 \sin( F_1 + F_2 ) \right] \label{eq:trimodal_dynamics_3D_2} \\
\dot{F}_2 &= \Delta\Omega_2 + K \left[ \gamma_1 \bar{r}_1 \bar{r}_2 \sin F_1 - (1-\gamma_1) \bar{r}_2 r_3 \sin F_2 - \gamma_1 \bar{r}_1 r_3 \sin( F_1 + F_2 ) \right], \label{eq:trimodal_dynamics_3D_3} 
\end{align}
\end{widetext}
where $F_m = f_{m+1}-f_m$ and $\Delta\Omega_m = \Omega_{m+1}-\Omega_m$. We see good agreement in the time series plots of $r_3$ for the 3D system (\ref{eq:trimodal_dynamics_3D_1})--(\ref{eq:trimodal_dynamics_3D_3}), shown as dashed black in Fig.~\ref{fig:trimodal_5D-vs-3D_time_series}) and for the full 6D collective coordinate system (\ref{eq:MCA_dynamics_infinite_3})--(\ref{eq:MCA_dynamics_infinite_4}), shown as solid gray in Fig.~\ref{fig:trimodal_5D-vs-3D_time_series}. In both models, $r_3$ experiences the same oscillations at the start and end, and both have a significant dip to $r_3\approx 0.4$ between $t=10$ and $t=30$. Furthermore, the Poincar\'{e} sections through the plane $F_1=0$, shown in the $(r_3,F_2)$-plane in Fig.~\ref{fig:trimodal_psection}, are similar for both models.

\begin{figure*}[tbp]
\centering
\includegraphics[width=\textwidth]{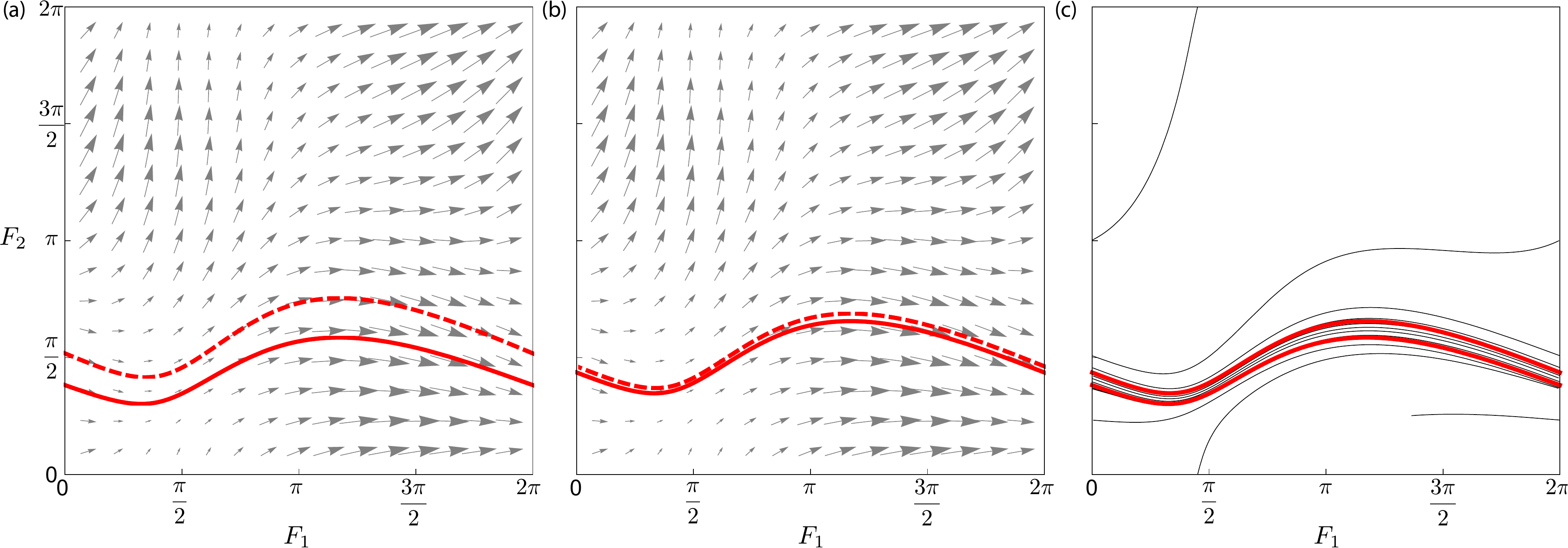}
\caption{(a,b)~The velocity field (\ref{eq:trimodal_dynamics_3D_2})--(\ref{eq:trimodal_dynamics_3D_3}) and stable (thick solid red) and unstable (thick dashed red) limit cycles in the $(F_1, F_2)$ plane for fixed $r_3$. (a)~$r_3=1$ and (b)~$r_3=0.981\approx r_c$, the critical value at which the limit cycles annihilate via a saddle-node bifurcation. (c)~The stable limit cycles from (a) and (b) are shown together with a trajectory of the full 6D collective coordinate model (\ref{eq:MCA_dynamics_infinite_3})--(\ref{eq:MCA_dynamics_infinite_4}), projected onto the $(F_1, F_2)$-plane (thin black). The tracer spends most of its time in the region between the stable limit cycles.}
\label{fig:trimodal_limit_cycles}
\end{figure*}

To explain the pronounced dips in $r_3$ in more detail, observe that for the time series of $r_3$, shown in Fig.~\ref{fig:trimodal_TS}(c), in between the sharp dips, $r_3$ exhibits small oscillations and a small negative trend. To explain this, let us assume that $r_3$ is constant, so the dynamics (\ref{eq:trimodal_dynamics_3D_1})--(\ref{eq:trimodal_dynamics_3D_3}) reduces to a 2D system for $F_1$ and $F_2$, given by (\ref{eq:trimodal_dynamics_3D_2})--(\ref{eq:trimodal_dynamics_3D_3}), with $r_3$ being a parameter. For $r_3 > r_c\approx 0.981$, this 2D system (\ref{eq:trimodal_dynamics_3D_2})--(\ref{eq:trimodal_dynamics_3D_3}) has one stable and one unstable limit cycle, as demonstrated in Fig.~\ref{fig:trimodal_limit_cycles}(a) for $r_3=1$ by the thick solid and dashed red curves, respectively. The gray arrows in Fig.~\ref{fig:trimodal_limit_cycles}(a) are the 2D velocity field. As $r_3$ decreases, the stable and unstable limit cycles move toward each other, as demonstrated in Fig.~\ref{fig:trimodal_limit_cycles}(b) for $r_3 = 0.981$. At $r_3=r_c$, the stable and unstable limit cycles annihilate via a saddle-node bifurcation, and the dynamics is topologically equivalent to quasiperiodic rotation on the torus. We observe in Fig.~\ref{fig:trimodal_limit_cycles}(c) that trajectories of the full 6D collective coordinate system (\ref{eq:MCA_dynamics_infinite_3})--(\ref{eq:MCA_dynamics_infinite_4}), projected onto the $F_1, F_2$ plane, follow curves that closely match the limit cycles corresponding to constant $r_3$. The tracer (whose trajectory is shown in thin black) slowly advances upward in between the lower limiting stable limit cycle corresponding to $r_3 =1$ (lower thick red curve), and the upper limiting stable limit cycle corresponding to $r_3 = r_c$ (upper thick red curve). This slow advance upward corresponds to the slow decay of $r_3$ in between the sharp dips.

Expanding further, starting at a time $t_0$ when $r_3\approx 1$, a tracer in the full 6D collective coordinate model (\ref{eq:MCA_dynamics_infinite_3})--(\ref{eq:MCA_dynamics_infinite_4}) will have a trajectory in the $F_1, F_2$ plane that is very similar to the limit cycle obtained from the assumption that $r_3$ is constant (equal to $r_3(t_0)$). However, while $r_3$ is approximately constant, it decreases slightly over one period of the limit cycle. We can approximate the decrease in $r_3$ by computing $\Delta r_3 = r_3(t_0 + T) - r_3(t_0)$, where $T=T(r_3(t_0))$ is the period of the stable limit cycle, denoted by $\mathcal{C}_{r_3}$, of the 2D system (\ref{eq:trimodal_dynamics_3D_2})--(\ref{eq:trimodal_dynamics_3D_3}) with $r_3=r_3(t_0)$ held constant. Here $r_3(t_0 + T)$ is found by integrating (\ref{eq:trimodal_dynamics_3D_1}) along the stable limit cycle $\mathcal{C}_{r_3}$. This is valid under the assumption that $r_{1,2,3}$ are all constant between $t=t_0$ and $t=t_0+T$. Note that the values $r_3(t_0+T)$ and $\Delta r_3$ are independent of the initial locations of $F_1, F_2$ on the limit cycle $\mathcal{C}_{r_3}$. We find that $\Delta r_3 < 0$ for all $r_3>r_c$, and so it is inevitable that $r_3$ will eventually reach the critical value, $r_c$, where the stable limit cycle bifurcates. 
\\

\begin{figure}[tbp]
\centering
\includegraphics[width=\columnwidth]{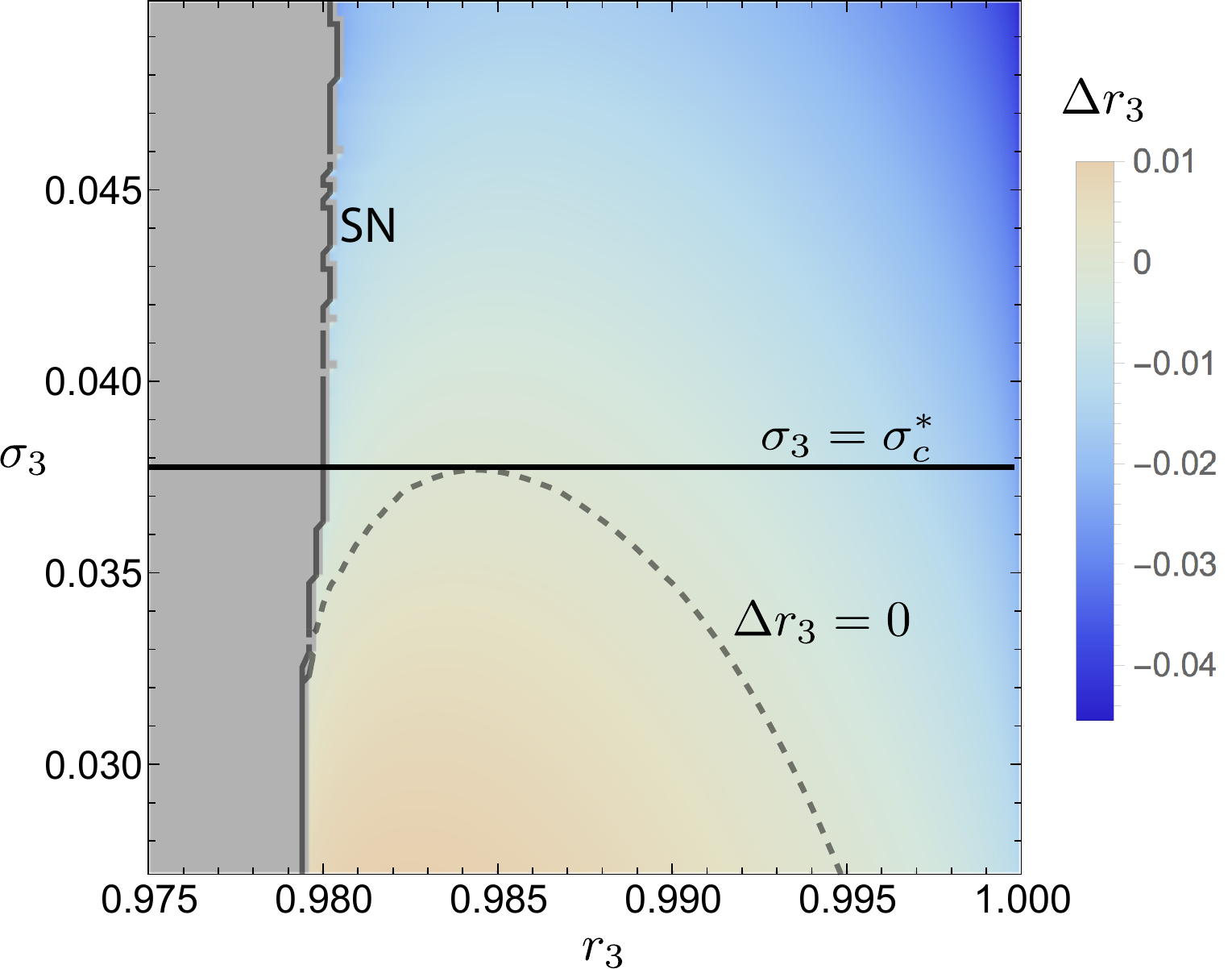}
\caption{The change in $r_3$ over one period of the associated limit cycle, $\Delta r_3$, over a range of $r_3$ and $\sigma_3$ values. Limit cycles do not exist in the gray region, to the left of the saddle-node (SN) bifurcation curve (solid gray). The curve $\Delta r_3 = 0$ (dashed black) indicates fixed points of the map $r_3 \mapsto r_3 + \Delta r_3$ for each value of $\sigma_3$. The line $\sigma_3=\sigma_c^*$ (solid black) separates chaotic cases ($\sigma_3>\sigma_c^*$) and non-chaotic cases ($\sigma_3<\sigma_c^*$). }
\label{fig:trimodal_dr3}
\end{figure}

The scenario of chaotic dynamics through intermittent desynchronization events is a robust phenomenon, occurring for a range of parameters of the natural frequency distribution (\ref{eq:multimodal_distribution}). We show this by investigating the effect of varying $\sigma_3$. As $\sigma_3$ decreases, we observe that the average time interval between dips in $r_3$ increases, and at a critical value of $\sigma_3=\sigma_c\approx 0.035$ the dips no longer occur. For $\sigma_3 < \sigma_c$, $r_3$ remains close to 1 for all time, and so the slow-fast splitting found in \S\ref{sec:time_scale_splitting} is valid, and the dynamics is non-chaotic. The value of $\sigma_c$ can be estimated using the collective coordinate system (\ref{eq:trimodal_dynamics_3D_1})--(\ref{eq:trimodal_dynamics_3D_3}). Consider $\Delta r_3$, the change in $r_3$ over the stable limit cycle that exists under the assumption that $r_3$ is constant. The distribution of $\Delta r_3$ across a range of $r_3$ and $\sigma_3$ values is shown in Fig.~\ref{fig:trimodal_dr3}. The turning point of the curve $\Delta r_3 = 0$ (dashed black in Fig.~\ref{fig:trimodal_dr3}) yields the approximation $\sigma_c^* = 0.038$ (solid black line in Fig.~\ref{fig:trimodal_dr3}) for $\sigma_c$. For $\sigma_3> \sigma_c^*$, $\Delta r_3$ is negative for all values of $r_3$ that have a stable limit cycle. Hence $r_3$ decreases after each period of the limit cycle until reaching the saddle-node bifurcation (solid gray curve in Fig.~\ref{fig:trimodal_dr3}). For $\sigma_3< \sigma_c^*$, the curve $\Delta r_3=0$ (dashed black in Fig.~\ref{fig:trimodal_dr3}) indicates the locations of fixed points of the map $r_3 \mapsto r_3 + \Delta r_3$, with the right-most fixed point being stable. The presence of these stable fixed points indicates a periodic solution of the three-dimensional system (\ref{eq:trimodal_dynamics_3D_1})--(\ref{eq:trimodal_dynamics_3D_3}). Therefore, $\sigma_c^*$ represents a bifurcation between periodic dynamics and intermittent desynchronization dynamics, i.e., it is an approximation for $\sigma_c$. Note that $\sigma_c^*=0.038$ slightly over-predicts $\sigma_c=0.035$, which is due to the inaccuracies that occur from making the assumption that $r_3$ is constant over the period of the limit cycle $\mathcal{C}_{r_3}$, when, as we have seen, it is both oscillating and slowly decreasing [cf. Fig.~\ref{fig:trimodal_snapshots}(a)]. A similar approach can be used to determine critical values at which chaos ceases to occur when other parameters in the natural frequency distribution are varied, such as the distance between peaks and the proportion of oscillators in each cluster.

We now explain why the transition into desynchronization occurs on a fast time-scale, as observed in Fig.~\ref{fig:trimodal_snapshots}(a), using the collective coordinate equations (\ref{eq:trimodal_dynamics_3D_1})--(\ref{eq:trimodal_dynamics_3D_3}) for the three-cluster interactions. The intercluster interaction term between the second and third cluster in $\dot{r}_3$ is $G(r_3) \cos F_2$, where $G(r_3)$ scales as $r_3^2 \sqrt{-2\log{r_3}}$ [see (\ref{eq:trimodal_dynamics_3D_1})], which is positive for $0<r_3<1$, and equal to zero at $r_3=0 $ and $r_3=1$. On the stable limit cycles, $F_2$ oscillates around $\pi/2$. Therefore, the interaction term is small while on the limit cycles, but when $r_3<r_c$, and the saddle-node bifurcation occurs, $F_2$ increases away from $\pi/2$, and so $\dot{r}_3$ becomes strongly negative, explaining the sharp decline of $r_3$. At the point where $F_2$ crosses $3\pi/2$, the sign of $\cos(F_2)$ changes, and so $\dot{r}_3$ becomes strongly positive, until $F_2$ once again approaches $\pi/2$, at which point $r_3$ once again becomes slow, and the system relaxes to a limit cycle corresponding to $r_3\approx 1$. This restarts the cycle of slow decay followed by a sharp decline and recovery.

We have established how chaos is generated through the delicate interaction of three clusters using the collective coordinate framework. As a summary, chaos occurs as a sensitivity between the entry and exit locations to the regular limit cycle zone. This sensitivity is shown by the infinite, fractal accumulation of folds in the zoomed in Poincar\'{e} section through the plane $F_1 = 0$ (cf. Fig.~\ref{fig:trimodal_psection}). The folds accumulate in the small region with $r_3\approx 1$ and $F_2\approx \pi/2$, corresponding to the regular limit cycle dynamics and slow, predictable decay of $r_3$. While we have shown that chaos is possible for trimodal natural frequency distributions, it is a rare phenomenon. In the process of finding the natural frequency distribution shown in Fig.~\ref{fig:trimodal_PDF}, we computed the maximal Lyapunov exponent for $5\times 10^4$ randomly drawn sets of natural frequency distribution parameters $(\bm{\Omega},\bm{\sigma},\bm{\gamma})$ and coupling strengths $K$ that produce synchronized clusters, and found that only $90$ cases were chaotic (with a positive Lyapunov exponent), i.e., only $0.18\%$.

\section{Two clusters: no chaos} \label{sec:two_clusters}

For two clusters, $M=2$, the thermodynamic limit collective coordinate equations (\ref{eq:MCA_dynamics_infinite_3})--(\ref{eq:MCA_dynamics_infinite_4}) become
\begin{align}
\dot{r}_1 &=-\sigma_1^2 \beta_1 r_1\left( 1 - K\beta_1 r_1 \left( \gamma_1 r_1 + \gamma_2 r_2 \cos F\right) \right) \label{eq:bimodal_dynamics_1}  \\
\dot{r}_2 &= -\sigma_2^2 \beta_2 r_2\left(1 - K\beta_2 r_2 \left( \gamma_2 r_2 + \gamma_1 r_1 \cos F \right) \right)  \\
\dot{F} &= \Delta \Omega - K r_1 r_2 \sin F, \label{eq:bimodal_dynamics_3}
\end{align}
where $F = f_2 - f_1$ is the phase difference of the two clusters, and $\Delta \Omega = \Omega_2 - \Omega_1$. Hence, it appears that there are three degrees of freedom, and chaos is theoretically possible.

We now show, using the collective coordinate approach, that chaos is not possible. In particular, phase chaos is not possible as it would reduce the dimension of the system to 1D with both $r_1$ and $r_2$ being constant. The case of intermittent desynchronization leads to decoupled 1D slow and 2D fast dynamics, excluding the possibility of chaos.

\begin{figure*}[tbp]
\centering
\includegraphics[width=\textwidth]{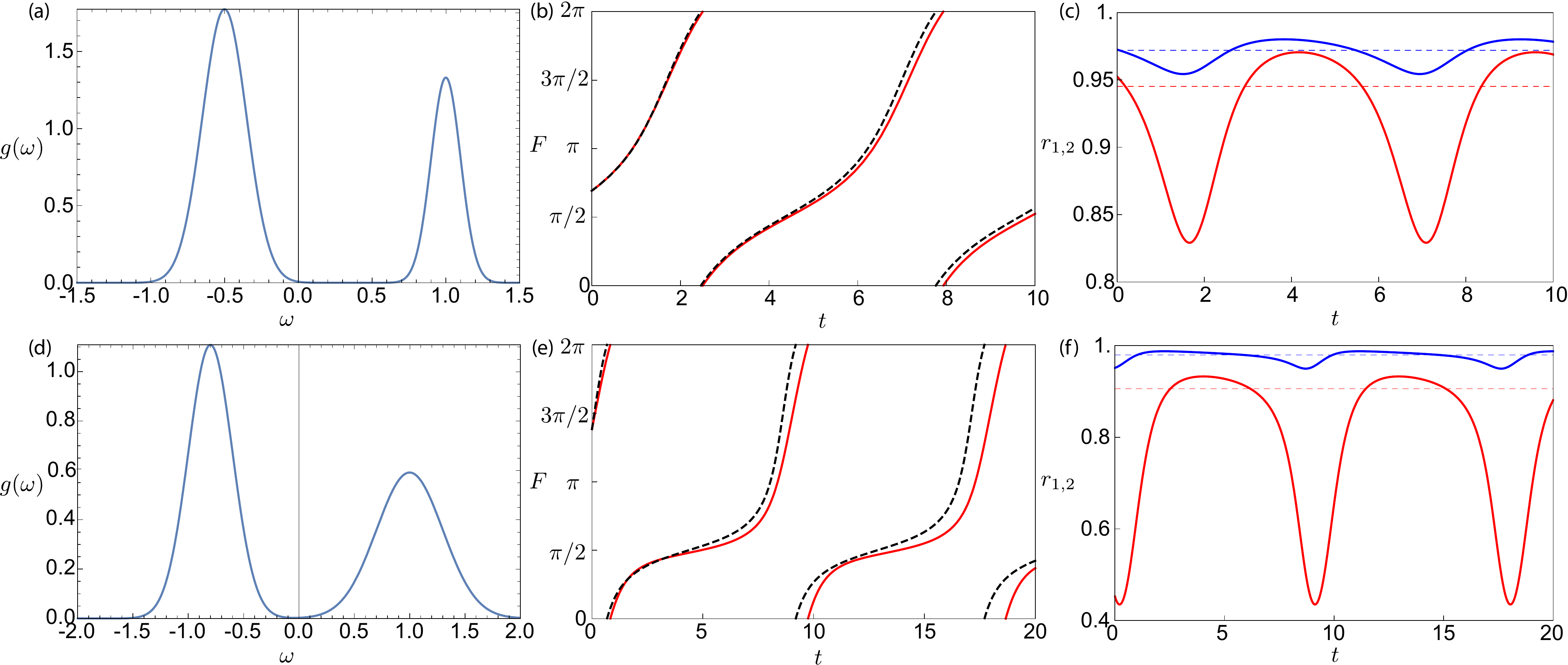}
\caption{Dynamics for two bimodal frequency distributions. (a,d)~Natural frequency distributions, $g(\omega)$. (a):~$(\sigma_1,\sigma_2)=(0.15,0.1)$, $(\Omega_1,\Omega_2) = (-0.5,1)$, $(\gamma_1,\gamma_2) = (2/3,1/3)$. (d):~$(\sigma_1,\sigma_2)=(0.2,0.3)$, $(\Omega_1,\Omega_2) = (-0.8,1)$, $(\gamma_1,\gamma_2) = (5/9,4/9)$. (b,e)~Time series of $F$ for the 3D collective coordinate model (\ref{eq:bimodal_dynamics_1})--(\ref{eq:bimodal_dynamics_3}) (solid red), and the function $F(t)$ given by (\ref{eq:bimodal_F_solution}) (dashed black). (b)~$K=1$, (e)~$K=1.85$. (c,f)~Time series of $r_1$ (solid blue, upper) and $r_2$ (solid red, lower). Also shown are the values of $r_1$ (dashed blue) and $r_2$ (dashed red) that are the stable solutions to the single cluster ansatz (\ref{eq:SCA_evolution_infinite}) for each cluster. These are also the stationary solutions of the time-averaged dynamics, assuming the time-scale splitting between $F$ (fast) and $r_{1,2}$ (slow) is valid.}
\label{fig:bimodal}
\end{figure*}

Let us begin with excluding the possibility of phase chaos. If the time scale splitting between $r_1,r_2$ (slow) and $F$ (fast) is valid, i.e., if the two clusters remain synchronized for all time, we can average the slow $r_1,r_2$ dynamics over the fast dynamics $F$. Assuming $r_1$ and $r_2$ are constant, the dynamics of $F$ can be solved analytically, with solution
\begin{equation} \label{eq:bimodal_F_solution}
F(t) = 2 \arctan \left( \frac{\kappa}{\Delta \Omega} + \frac{B}{\Delta \Omega} \tan \left( \frac{B}{2}  t + C \right) \right)
\end{equation}
where $\kappa = K r_1 r_2$, $B= \sqrt{\Delta \Omega^2 - \kappa^2}$, $C =  \arctan \left( \frac{ -\kappa + \Delta \Omega \tan \frac{F_0}{2}}{B} \right)$, and $F(0)=F_0$. Note that $F$ is a periodic function, with period $T=2\pi/B$. For the bimodal frequency distribution shown in Fig.~\ref{fig:bimodal}(a), where the peaks have very little overlap, the approximate solution (\ref{eq:bimodal_F_solution}) [dashed black in Fig.~\ref{fig:bimodal}(b)] for the phase difference closely matches the time series of $F$ of the collective coordinate model (\ref{eq:bimodal_dynamics_1})--(\ref{eq:bimodal_dynamics_3}) [solid red in Fig.~\ref{fig:bimodal}(b)]. 

Furthermore, since $F(t)$ ranges from $0$ to $\pi$, we can choose, without loss of generality, our starting time such that $F_0=\pi/2$, and so $C=\arctan \frac{\Delta\Omega - \kappa}{B}$. It can be shown that
\begin{equation} \nonumber
F(t) = \frac{\pi}{2} + 2 \arctan \left[ \frac{\Delta\Omega-\kappa}{B}\tan\left(\frac{B}{2}t\right) \right],
\end{equation}
which implies that
\begin{equation} \nonumber
\cos F(t) = -\sin\left( 2 \arctan \left[ \frac{\Delta\Omega-\kappa}{B}\tan\left(\frac{B}{2}t\right) \right] \right).
\end{equation}
Therefore, $\cos F(t)$ is an odd periodic function,  and so its average over one period, $\langle \cos F(t) \rangle$, is zero. This means the dynamics of the time-averaged variables $\bar{r}_1, \bar{r}_2$ becomes decoupled from one another. The dynamics for each cluster is equivalent to the single cluster ansatz equation (\ref{eq:SCA_evolution_infinite}), with $K$ replaced by $K \gamma_i$ for $i=1,2$. Hence, $r_1(t)$ [solid blue curve in Fig.~\ref{fig:bimodal}(c)] and $r_2(t)$ [solid red curve in Fig.~\ref{fig:bimodal}(c)] oscillate around the stable equilibria, $r_1^*$ and $r_2^*$ (dashed blue and red resp.), obtained from the respective single cluster ansatz equations, and phase chaos cannot occur.

Now we go on to exclude the case that one cluster intermittently desynchronizes, like in the three cluster case discussed in the previous section. This occurs for the natural frequency distribution shown in Fig.~\ref{fig:bimodal}(d), where the second cluster intermittently desynchronizes, approaching $r_2\approx 0.4$, as shown by the solid red curve in Fig.~\ref{fig:bimodal}(f). In this case, the dynamics of $r_1$, which remains close to 1 for all time, is slower than $r_2$ and $F$. This is confirmed in Fig.~\ref{fig:bimodal}(e), where it is shown that the time evolution of $F$ given by numerical simulation of (\ref{eq:bimodal_dynamics_1})--(\ref{eq:bimodal_dynamics_3}) (solid red) is not well approximated by the function (\ref{eq:bimodal_F_solution}) (dashed black), which assumes perfect time-scale splitting. Therefore, we may not assume time-scale separation between $r_2$ and $F$. We have an effective 2D fast system for $r_2$ and $F$. This 2D system has a stable limit cycle in cases with two clusters that do not globally synchronize. In turn, the dynamics of $r_1$ cannot be chaotic, since the time-averaged dynamics is a 1D system with time-periodic forcing. 

The only other possibility is that both clusters intermittently desynchronize. However, since it is the intercluster terms in $\dot{r}_1$ and $\dot{r}_2$ that drive the push away from the single cluster ansatz equilibrium, and both intercluster terms are multiples of $r_1 r_2\cos F$, it follows that $r_1$ cannot rapidly decay without $r_2$ also rapidly decaying. If one, say $r_1$, decays faster than the second, $r_2$, then it will asymptote toward $r_1=0$, and so it has no effect on the second cluster. The second cluster is then governed by the single cluster ansatz, and will either approach the stable synchronized state, or approach $r_2=0$, depending on whether $r_2$ crosses the unstable fixed point of the single cluster ansatz equation while the first cluster is desynchronizing. In either case, the dynamics is regular, and stationary in the long run. If both $r_1$ and $r_2$ decay at the same rate, then the system possesses a symmetry, which further reduces the effective dimension, excluding the possibility of chaos. 
\\

The above discussion used the thermodynamic limit. In finite size networks, however, chaos can occur for bimodal natural frequency distributions. This occurs due to sampling effects. In our numerical simulations of finite size networks, we found that it is typical that when chaos occurs, a small group of oscillators, with natural frequencies at one or the other extreme of the distribution (i.e., very high or very low), do not synchronize with the other oscillators corresponding to the same peak in the natural frequency distribution. This group of ``rogues'' may either constitute a set of incoherent oscillators or another small cluster. In either case, the system must be considered as having more than two clusters, which agrees with our results obtained in \S\ref{sec:four_clusters} and \S\ref{sec:three_clusters}. We find fewer chaotic cases as the number of oscillators increases, which confirms that the issue is a finite-size effect. It is important to note that we have found no bimodal cases with finite $N$ that are chaotic and do not have unsynchronized rogue oscillators.

\section{Summary and outlook} \label{sec:conclusions}

\subsection{Summary}

Employing detailed numerical simulations guided by analytical results from a collective coordinate reduction we have established necessary conditions for collective chaos in the Kuramoto model with multimodal natural frequency distributions. We have shown that phase chaos can occur provided there are at least four peaks in the natural frequency distribution. This is due to a time-scale splitting between slow intracluster collective coordinates and fast intercluster collective coordinates, which reduces the Kuramoto model to $M-1$ active degrees of freedom, where $M$ is the number of peaks in the natural frequency distribution.  

For three peaks in the natural frequency distribution, we have shown that chaos can occur via intermittent desynchronization of clusters. When a cluster desynchronizes, its intracluster collective coordinate becomes fast, resulting in an additional active degree of freedom. Through the slow-fast splitting, the collective coordinate description has allowed us to study the intricate dynamics of intermittent desynchronization, and show that it is a robust phenomenon.

For two peaks in the natural frequency distribution, the collective coordinate description has allowed us to rule out the possibility of chaos.

We have shown that for both phase chaos and chaos via intermittent desynchronization, the reduced collective coordinate description can be used to quantitatively predict the leading Lyapunov exponent, and, hence, regions of the parameter space where chaos occurs.

However, it is important to note that these results are primarily for the thermodynamic limit. For finite size networks, even bimodal natural frequency distributions can be chaotic. In those cases, there are rogue oscillators that do not synchronize with the rest of their cluster. These rogues can be treated as separate clusters, each of which requiring its own additional collective coordinate, increasing the number of active degrees of freedom, and opening up the possibility of chaos.

\subsection{Outlook}

\begin{figure}[tbp]
\centering
\includegraphics[width=\columnwidth]{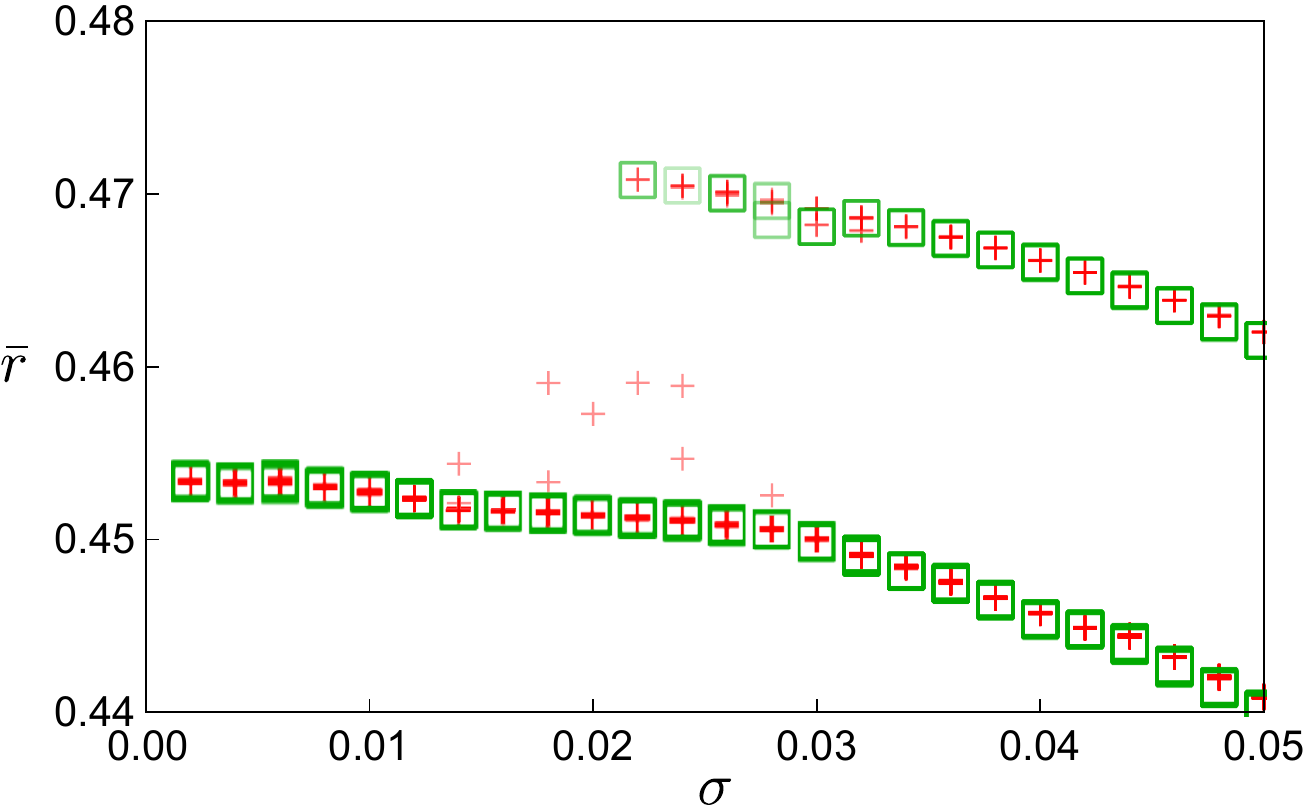}
\caption{Multistability of the order parameter $\bar{r}$ for $K=0.95$ and multimodal natural frequency distributions with four peaks, as in Fig.~\ref{fig:quadmodal_distro}, such that each peak has variance $\sigma^2$. Shown are results for the full Kuramoto model (\ref{eq:full_KM}) with $N=100$ oscillators (green squares) and the collective coordinate model (\ref{eq:MCA_dynamics_finite_1})--(\ref{eq:MCA_dynamics_finite_2}) with $N=100$ (red +'s). For each model, 100 random initial conditions are seeded to determine regions of multistability.}
\label{fig:multistability}
\end{figure}

In our numerical simulations, we have observed regions in the parameter space of multimodal natural frequency distributions with four peaks that exhibit multistability, including natural frequency distributions that yield both strange attractors and limit cycles, depending on the initial condition. For example, Fig.~\ref{fig:multistability} shows that for $K=0.95$ and multimodal distributions like Fig.~\ref{fig:quadmodal_distro}, a second stable branch exists for $\sigma>0.022$ for the full Kuramoto model (\ref{eq:full_KM}) with $N=100$ oscillators (green squares). This multistability is well reproduced by the collective coordinate model (\ref{eq:MCA_dynamics_finite_1})--(\ref{eq:MCA_dynamics_finite_2}) with $N=100$ (red +'s) and by the collective coordinate model in the thermodynamic limit (\ref{eq:MCA_dynamics_infinite_1})--(\ref{eq:MCA_dynamics_infinite_2}) (not shown). On the lower branch, the dynamics is periodic, and has the property that $r_1(t) = r_4(t+T/2)$ and $r_2(t) = r_3(t+T/2)$, where $T$ is the period of the system. On the upper stable branch there is no such relation between the cluster order parameters. Further study is required to understand this phenomenon, and the bifurcations that control it. Since the reduced collective coordinate models are more analytically tractable than the full Kuramoto model and accurately predict the existence of multistability, they may be used to provide deeper insight into this phenomenon. 

Here we have considered all-to-all networks with synchronized clusters that result from distinct peaks in the natural frequency distribution. However, synchronized clusters can also occur due to the network topology. Future studies should consider whether topological clusters can yield chaos. Furthermore, chaos could result from a combination of frequency clustering and topological clustering. For example, a bimodal natural frequency distribution and a network with two clusters, which would result in four synchronized clusters of oscillators and the three degrees of freedom required for chaos.

\begin{acknowledgments}
We wish to acknowledge support from the Australian Research Council, Grant No. DP180101991.
\end{acknowledgments}


%

\end{document}